%% file: paper.tex
\documentclass[sigconf]{acmart}

\renewcommand\footnotetextcopyrightpermission[1]{} 

\usepackage{booktabs} 

\newcommand{\pseudosection}[1]{\vspace{0.5\baselineskip} \noindent {\bf #1}}

\setcopyright{none}

\begin{document}
\title{A Framework for Exploring and Evaluating Mechanics in Human Computation Games}

\author{Kristin Siu}
\affiliation{
	\department{School of Interactive Computing}
	\institution{Georgia Institute of Technology}}
\email{kasiu@gatech.edu}

\author{Alexander Zook}
\affiliation{
	\department{School of Interactive Computing}
	\institution{Georgia Institute of Technology}}
\email{a.zook@gatech.edu}

\author{Mark O. Riedl}
\affiliation{
	\department{School of Interactive Computing}
	\institution{Georgia Institute of Technology}}
\email{riedl@cc.gatech.edu}

\renewcommand{\shortauthors}{K. Siu, A. Zook, and M. Riedl}

\begin{abstract}
Human computation games (HCGs) are a crowdsourcing approach to solving computationally-intractable tasks using games.
In this paper, we describe the need for generalizable HCG design knowledge that accommodates the needs of both players and tasks.
We propose a formal representation of the \emph{mechanics} in HCGs, providing a structural breakdown to visualize, compare, and explore the space of HCG mechanics.
We present a methodology based on small-scale design experiments using fixed tasks while varying game elements to observe effects on both the \emph{player experience} and the human computation \emph{task completion}.
Finally we discuss applications of our framework using comparisons of prior HCGs and recent design experiments.
Ultimately, we wish to enable easier exploration and development of HCGs, helping these games provide meaningful player experiences while solving difficult problems.
\end{abstract}

%
%



\maketitle

\input{introduction}
\input{related-work}
\input{formalization}
\input{methodology}
\input{examples}

\input{conclusions}

\begin{acks}
This material is based upon work supported by the National Science Foundation under Grant No. 1525967.
Any opinions, findings, and conclusions or recommendations expressed in this material are those of the author(s) and do not necessarily reflect the views of the National Science Foundation.
\end{acks}

\bibliographystyle{ACM-Reference-Format}
\bibliography{bibliography}

\end{document}

%% file: introduction.tex
\section{Introduction}
\noindent Games are everywhere.
Games accompany people out on their mobile phones and await them back at home on their entertainment systems.
Games are being integrated into educational curricula, embedded in wearable devices, and used in professional training simulations.
For \emph{human computation}, which harnesses the computational potential of the human crowd, this increasingly diverse audience of players represents new opportunities for computation. 
\emph{Human computation games} (HCGs) are games that ask players to help solve complex, computationally-intractable tasks or provide data through gameplay.
These games---also known as \emph{Games With a Purpose} (GWAPs), scientific discovery games, and citizen science games---have been used to solve a wide variety of problems such as image labeling, protein folding, and data collection.

Compared with mainstream games for entertainment, one hurdle compounding HCG development is that these games suffer the design problem of serving two different goals.
On the one hand, an HCG must provide a sufficiently-engaging experience for its \emph{players}.
On the other hand, an HCG must enable players to successfully complete the underlying human computation \emph{task}.
Balancing these two goals is difficult and often results in conflicting design decisions.
To compound this dilemma, very little design knowledge exists beyond a small number of simple patterns from examples or takeaways from successful games (e.g.,~\cite{cooper2010:sci-disc-design, vonahn2008:gwap-design}).
This lack of knowledge is intimidating to task providers and HCG developers, who might find it difficult to justify the risk of an expensive development process when there are no guarantees that a game will be successful at a desired computational task.
As a result, most HCGs to date are built around specific kinds of templates, leaving the space of possible HCG designs limited and relatively unexplored.

To facilitate broader adoption and ease of game development, human computation game design needs the tools and frameworks to study and communicate about these games in a consistent manner.
We need to understand precisely what game elements make certain HCGs successful, that is both effective at engaging players and solving tasks.
Building up comprehensive, reusable design knowledge for HCGs that addresses both players and tasks enables us to construct formal representations for tasks, audiences, and game elements.
A common language and structure for HCGs would allow us to talk about and explore the space of possible HCG designs, thus broadening the diversity of HCGs as a whole.

Additionally, this knowledge could also be used to build formal models that can predict the effect of tasks, audiences, and game elements on both aspects of the player experience---engagement and retention---and the completed task metrics---data quality, volume, and diversity.
All of these affordances would help to make HCGs more successful interfaces for solving new tasks while still able to engage new player audiences with changing player preferences.

In this paper, we contend that reusable design knowledge for human computation games is necessary and should take the needs of both players and tasks into account.
We introduce a formal representation of HCG mechanics that provides us with a common vocabulary and structure to visualize, compare, and explore the space of game mechanics in HCGs.
We advocate a methodology for building up HCG design knowledge, which uses small-scale, controlled design experiments on tasks with known solutions to understand how variations of game elements affect the player experience and the completion of human computation tasks.
Finally, we illustrate how our representation, combined with this methodology, enables the comparative study of existing HCGs and the exploration of HCG mechanics through novel designs.

%% file: related-work.tex
\section{Background}
Human computation is the process of using people to solve computational problems as an alternative for current algorithms~\cite{law2011:hcomp-book}.
Such problems are often broken into smaller tasks, which are completed by crowdsourced workers, and then aggregated into larger solutions.
This paradigm has been used for tasks including classifying objects, ranking items, summarizing texts, and iterative document editing and improvement~\cite{law2011:hcomp-book}.
Some human computation or crowdsourcing tasks rely on volunteer efforts, where workers are motivated by participation in scientific processes or altruistic goodwill.
More common are online interfaces such as Amazon Mechanical Turk and CrowdFlower, which give task providers a platform to recruit workers, distribute tasks, and provide monetary compensation.

Human computation games have been developed as an alternative to these monetary incentive systems, utilizing game mechanics to enable task completion and providing players an engaging gameplay experience, in addition or as an alternative to financial compensation.
The original Game With a Purpose, the \emph{ESP Game}, addressed the problem of labeling images~\cite{vonahn2004:esp}, but the breadth of human-computed tasks has greatly diversified since.
HCGs have been used to annotate or classify other kinds of information, from music~\cite{barrington2012:herdit, law2009:tagatune} to galaxies~\cite{lintott2008:galaxyzoo}, to relational information (e.g., ontology construction)~\cite{krause2010:ontogalaxy, siorpaes2008:ontogame}, to protein function recognition~\cite{peplow2016:eve-project-discovery}.
Other HCGs have leveraged human players as alternatives to optimization functions in order to manipulate and refine existing input information.
Common tasks of this type are often ``scientific discovery'' problems such as protein~\cite{cooper2010:foldit} and RNA folding~\cite{lee2014:eterna}, DNA multiple sequence alignment~\cite{kawrykow2012:phylo}, software verification~\cite{dietl2012:verigames,logas2014:xylem}, and mapping dataflow diagrams onto hardware architectures~\cite{mehta2013:untangled-mapping-game}.
Additionally, HCGs have used crowdsourced audiences to collect or generate new information, such as creative content or datasets for machine-learning algorithms.
These tasks include photo collection~\cite{tuite2011:photocity, tuite2015:pointcraft}, location tagging~\cite{bell2009:eyespy, frazier2014:proactive-sensing}, and commonsense knowledge acquisition~\cite{kuo2009:virtualpetgame}.
Comprehensive taxonomies~\cite{krause2011:hcg-survey, pe-than2013:hcg-typology} detail a wide breadth of HCGs and the tasks they have tackled.

Human computation game design has been primarily guided by examples of successful games.
These include von Ahn and Dabbish's templates for classification and labeling tasks~\cite{vonahn2008:gwap-design} and the design anecdotes of \emph{Foldit}~\cite{cooper2010:sci-disc-design} rather than systematic study of HCG elements.
As a result, we do not understand what specific elements of these particular design choices work and how to appropriately generalize them or consider alternatives.
Confounding this issue is the fact that HCG research remains divided on how game elements, in particular game \emph{mechanics}, can ensure both successful completion of tasks and engaging player experiences.
Some argue that HCG game mechanics should be \emph{isomorphic} or \emph{non-orthogonal} to the underlying task, that is game mechanics should map to the process of solving the underlying task~\cite{jamieson2012:hcomp-games,tuite2014:gwap-problem}.
Others argue that incorporation or adaptation of game mechanics from successful digital games designed for entertainment can leverage player familiarity with existing games and keep them more engaged~\cite{krause2010:ontogalaxy}.
This ongoing debate highlights the challenge of designing HCGs that optimize for both a positive player experience and a successfully-completed human computation task.

Prior research has demonstrated that qualitative and quantitative research methods can be used to study game design including methods from game usability~\cite{isbister2008:game-usability}, game analytics~\cite{seifel-nasr2013:game-analytics-book}, and visual analysis~\cite{wallner2013:gameplay-viz-analysis}.
Controlled studies have proven successful for understanding the influence of general game design elements, particularly in dual-purpose domains such as educational games.
Such studies have investigated game aspects including difficulty~\cite{lomas2013:opt-edugame-challenge, whitlock2014:flow}, controls~\cite{lankes2014:player-gaze, mcewan2014:natural-mapping}, and tutorials~\cite{andersen2011:harmful-objectives, andersen2012:tutorials-impact, liu2014:knowl-vs-learning-bandit, shannon2013:tutorial-practices}.

General crowdsourcing and human computation research has worked to formalize the design and presentation of crowdsourcing tasks~\cite{law2011:hcomp-book}.
Controlled studies in this domain focus on evaluating how design variations affect worker efficiency and task completion, examples of which include the kinds of tasks~\cite{doroudi2016:towards-learning-science-for-crowdsourcing} and the means of motivating workers~\cite{law2016:curiosity-cat-crowdsourcing}.
How these might apply to human computation games in a way that addresses the necessity to engage players remains unexplored.

It is only recently that researchers have conducted similar controlled studies on specific game elements of HCGs that jointly address aspects of the player experience and the completion of the human computation task~\cite{goh2011:fight-unite}, and advocated for their use~\cite{siu2014:gwap}.
We discuss these, along with other relevant examples in subsequent sections.
Combined with formal crowdsourcing research, we posit that these approaches can enable a formal study of HCG design, allowing HCGs to more effectively address a broader range of tasks in ways more satisfying to players.

%% file: formalization.tex
\section{Formalizing HCG Mechanics}
\label{sec:formalization}

We propose a formal representation of the mechanics of human computation games.
This representation serves three core functions: 
\begin{enumerate}
\item Provides a common vocabulary and visual organization of HCG elements
\item Enables formal comparison of existing HCGs to understand the space of HCG designs and their consequences
\item Facilitates the formulation of controlled design experiments of HCG elements to build further, generalizable knowledge of HCG design
\end{enumerate}

\begin{figure}[tb]
\centering
\includegraphics[width=0.5\textwidth]{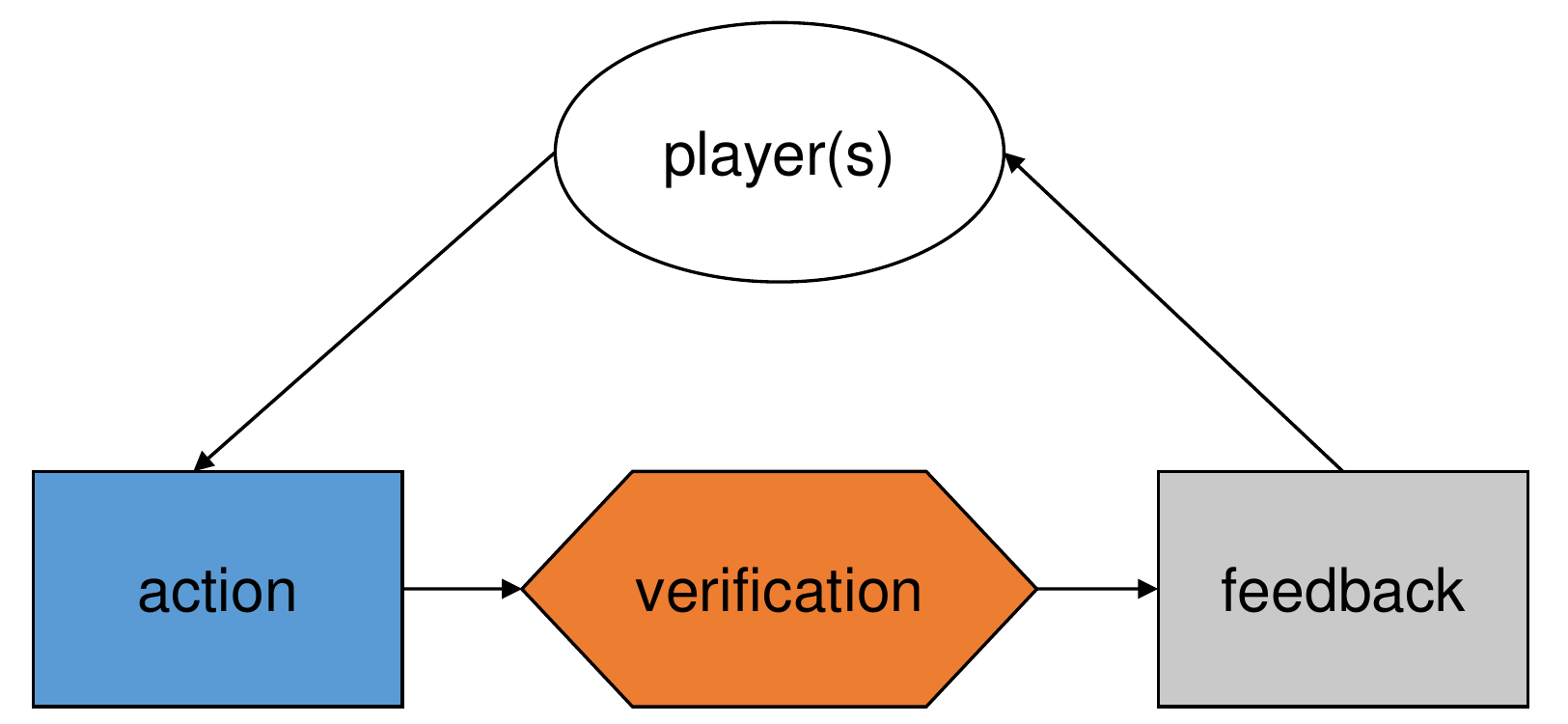}
\caption{Breakdown of HCG mechanics. Players provide inputs to take actions (shown in blue), which are verified (shown in orange), and receive feedback (shown in gray) from the game. Solid lines represent transitions through the gameplay loop.}
\label{fig:hcg-struct}
\vspace{-1.0\baselineskip}
\end{figure}

\noindent We specifically formalize human computation \emph{game mechanics}---the rules that define how a player can interact with the game systems---leaving other elements of HCG designs to future work.
We divide HCG game mechanics into three types: \emph{action} mechanics, \emph{verification} mechanics, and \emph{feedback} mechanics. 
As shown in \autoref{fig:hcg-struct}, this breakdown reflects the core gameplay loop of most HCGs.
HCGs begin with players taking in-game actions, then compare task-relevant input from these actions through verification mechanisms, and finally use verification output to provide feedback or reward for players.

\begin{figure*}[tb]
\centering
\includegraphics[width=\textwidth]{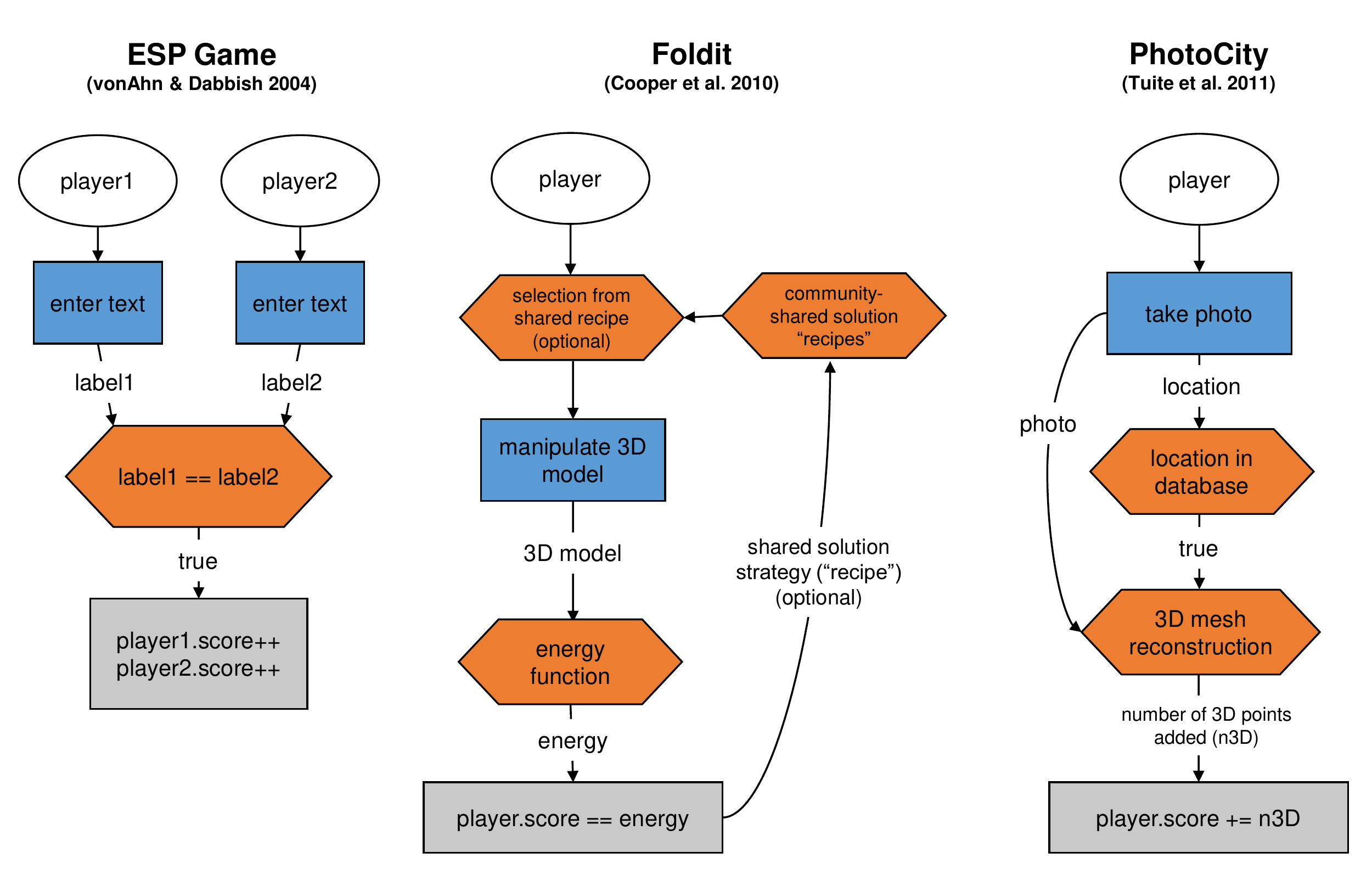}
\caption{Examples of HCGs~\cite{vonahn2004:esp,cooper2010:foldit,tuite2011:photocity} subdivided into \emph{action}, \emph{verification}, and \emph{feedback} mechanics. Arrows from feedback to players have been omitted for clarity.}
\label{fig:hcg-examples}
\end{figure*}

We now define and describe these three sets of mechanics in detail, illustrated using three successful HCGs spanning different tasks: the original \emph{ESP Game}~\cite{vonahn2004:esp}, \emph{Foldit}~\cite{cooper2010:foldit}, and \emph{PhotoCity}~\cite{tuite2011:photocity}.
Figure \ref{fig:hcg-examples} shows the mechanical breakdown of these games into \emph{action}, \emph{verification}, and \emph{feedback} mechanics.
In addition to these three examples, we also highlight notable instances of other HCGs that have explored these subsets of mechanics and discuss aspects of these mechanics that merit further exploration.

\subsection{Action Mechanics}
\label{sec:action-mechanics}
\emph{Action mechanics} are the interface for players to complete a human computation task through in-game actions or gameplay.
These mechanics align with the process of solving the human computation task, often asking players to utilize skills necessary for solving the task during play.
Such mechanics may be as simple as entering text input or as complicated as piloting a space ship in a virtual environment, and tend to vary based on the nature of the task. 

\pseudosection{Examples.}
In the \emph{ESP Game}, players provide labels through text entry to solve the task of labeling given images.
In \emph{Foldit}, players are given a variety of spatial actions, such as handling or rotating components of a protein structure, to solve the task of ``folding'' a given protein into a minimal energy configuration.
In \emph{PhotoCity}, players navigate to a desired location and take pictures using their camera phones, which are later uploaded to a database and used to construct a 3D representation of the buildings in that location.

\pseudosection{Discussion.}
In general, the \emph{action} mechanics in human computation games have been closely aligned with the process of solving the human computation task.
The three examples in Figure \ref{fig:hcg-examples} adhere to this alignment of mechanics and task completion, suggesting that the mechanics of these games were designed first and foremost with the task in mind, as opposed to adapting the mechanics of existing games for entertainment.

Actual explorations of adapting mechanics from mainstream or popular digital games do exist for HCGs, but are few.
For example, the game \emph{OnToGalaxy}~\cite{krause2010:ontogalaxy} converts an analogous classification task of that previously seen in games such as \emph{Ontogame}~\cite{siorpaes2008:ontogame} to a space shooter game akin to the classic arcade game \emph{Asteroids}.
Unfortunately no comparisons were conducted between these games; while \emph{OnToGalaxy} was compared against the \emph{ESP Game}, differences in tasks---not to mention game mechanics---make direct comparison and generalization difficult.
Meanwhile, the game \emph{Gwario}~\cite{siu2017:gwario} was used in an experiment to explore singleplayer and multiplayer mechanics in HCGs, turning a classification task from prior work~\cite{siu2014:gwap} into a platformer game resembling the classic action game \emph{Super Mario Bros.}.
To complement this study of game mechanic variations, the authors also solicited opinions from HCG developers and researchers, asking directly if mechanics from mainstream digital games should be adapted into HCGs.
Diverging from prior HCG design theories~\cite{jamieson2012:hcomp-games,tuite2014:gwap-problem}, these experts were positive towards such adaptation, though not without cautioning that such mechanics ought to be selected carefully to avoid compromising the task results.

One of the major concerns surrounding human computation games is that when compared against other games designed purely for entertainment, HCGs might be perceived as shallow~\cite{tuite2014:gwap-problem} and therefore risk failing to attract players.
Adaptation of familiar or successful mechanics from digital games might serve to address this issue.
Given that \emph{action} mechanics are often the first mechanics that players encounter in a game, an important question is: can actions be used to engage players and attract certain audiences?
Players accustomed to certain kinds of (\emph{action}) mechanics may find some games easier to interact with, easing entry into the game and facilitating early successes.
Conversely, familiar mechanics may raise expectations about the game that may or may not be met given the constraints of solving the task alongside expected gameplay.
For example, an HCG adopting \emph{action} mechanics from a successful game may negatively impact players' experiences if the design quality of the HCG did not match that of the popular game or series.

\subsection{Verification Mechanics}
\label{sec:verification-mechanics}
\emph{Verification mechanics} combine the output of player actions to compute task-relevant outcomes.
These mechanics can support task completion outcomes including the quality, volume, diversity, and the rate at which the data are acquired.

\pseudosection{Examples.}
For many human computation tasks, consensus on player input often serves as verification.
The \emph{ESP Game} (and many of the games inspired by its structure) verify using an online agreement check that filters correct answers from incorrect answers using agreement between players (Figure \ref{fig:hcg-examples}).
The \emph{ESP Game} later added ``taboo word'' mechanics to promote data diversity through banning words once consensus on existing data was reached.

By contrast, both \emph{Foldit} and \emph{PhotoCity}, which emphasize data manipulation and collection, handle verification through a task-based evaluation function.
\emph{Foldit}'s protein configuration energy function determines the quality of player solutions.
In \emph{PhotoCity}, the game does not explicitly evaluate the provided photos; photos are instead processed on an offline server and then player feedback is based on the resulting alterations to a constructed 3D mesh of the world.

\emph{Foldit} also makes use of social mechanics, such as allowing players to share solution procedures (called ``recipes'') through its community interfaces, as an additional (but optional) instance of verification~\cite{cooper2011:foldit-recipes}.
Players can utilize existing recipes uploaded by other players as a starting point for solving tasks, thus validating and iterating on pre-existing, partial solution strategies, and can also rate their utility.

\pseudosection{Discussion.}
Verification requirements for a task impact design decisions related to the number of players required to play the game and how they might interact with each other.
Whether or not verification is handled within the human computation game or as an offline process may impact if a game uses singleplayer mechanics, multiplayer mechanics, or both.
Many human computation tasks rely on consensus, and multiple players permits synchronous consensus, as shown in the \emph{ESP Game}.
Other tasks, which can be evaluated using an objective function or comparison against existing data, may permit asynchronous (singleplayer) play, such as \emph{Foldit} and \emph{PhotoCity}.
Synchronous verification often requires multiple players to be playing at the same time, necessitating a fast verification step for real-time feedback.
Note that simulated players can be used to re-verify old solutions (e.g., utilized in the \emph{ESP Game}), especially where multiple players are required, but are not necessarily available concurrently.

As with \emph{action} mechanics, few design experiments have tested alternative verification mechanics in HCGs.
The game \emph{KissKissBan}~\cite{ho2009:kiss-kiss-ban} modified the original structure of the \emph{ESP Game} to promote data diversity (i.e., a broader set of image labels), by adding a third player to ``ban'' common words.
Compared with the \emph{ESP Game's} eventual use of taboo words, \emph{KissKissBan} demonstrates an alternative \emph{verification} mechanic that players found engaging.

In most human computation games (i.e., those which rely on synchronous consensus), players are commonly forbidden from direct communication with each other (often implemented through anonymous pairing with no means of determining the identity of the other player).
Very few games (e.g., \emph{Foldit}) allow players to communicate through channels of communication such as game forums or community interfaces.
This is meant to minimize collusion between players, which may lead to players providing deliberately incorrect answers while still succeeding at the game.
The game \emph{Gwario}~\cite{siu2017:gwario} was used to compare singleplayer and co-located multiplayer, which allowed players to communicate directly during side-by-side play.
Study results found no negative impact from direct communication; collusion was instead found to be the strongest predictor of high task accuracy. 
This suggests that direct communication may be permissible for certain tasks, audiences, and kinds of games.

Overall, we believe that \emph{verification} mechanics are ripe with questions for future exploration.
How do different ways of determining agreement influence task results?
Would different mechanisms have downstream consequences for the feedback players can receive, for example, highlighting flaws in player-provided inputs?

\subsection{Feedback Mechanics}
\label{sec:feedback-mechanics}
\emph{Feedback mechanics} provide players with information or digital artifacts based on the results of player actions in terms of partial or full task completion.
These mechanics commonly encompass gameplay elements such as rewards and scoring, and can also be mapped to evaluation metrics for the underlying task, thus allowing both researchers and designers to assess player performance at both the completion of the task and progression through the in-game experience.

\pseudosection{Examples.}
For all of the games shown in Figure \ref{fig:hcg-examples}, players receive feedback in the form of a score.
However, the scale of the scoring mechanics themselves are unique to the tasks performed.
The \emph{ESP Game} rewards players with points for agreement on an image label.
By contrast, \emph{Foldit} rewards players with points for minimizing an energy function describing the protein structure.
\emph{PhotoCity} rewards players for the number of points their photo choices add to the reconstructed 3D mesh.
These examples are all similar in that the feedback ``currency'' is nominal---points contributing to a numerical score---but vary in what players are rewarded for.

\pseudosection{Discussion.}
Aspects of \emph{feedback} mechanics are some of the most well-explored design elements in human computation games.
These mechanics are synonymous with in-game rewards, which some consider comparable to monetary compensation in other crowdsourcing systems.
So what should rewards to the player represent: player actions in the game, player performance at the task (as evaluated by \emph{verification} mechanics), or a combination of both?

Traditionally, positive \emph{feedback} in human computation games is designed to encourage participation in the crowdsourcing process, as well as correctness of task completion.
For example, players may receive rewards for first completing a task and, if such verification is available, additional rewards for completing it correctly.
This kind of positive feedback often rewards collaborative player behavior, which is reasonable given that crowdsourcing is an aggregate process.
However, reward systems in games, such as points and leaderboards, are often designed to afford and encourage competitive player behavior.
A variety of research has examined in-game rewards for collaborative versus competitive behaviors, specifically in the context of HCGs where competition may be discouraged as a distraction from the task (i.e., players will optimize their behaviors to outperform others, rather than to complete the task sufficiently).
Both the game \emph{KissKissBan}~\cite{ho2009:kiss-kiss-ban} and a study by Goh et al.~\cite{goh2011:fight-unite} examine collaboration and competition in the context of the \emph{ESP Game}.
\emph{KissKissBan} explored the introduction of competitive mechanics by rewarding a third player for antagonistic behavior that would encourage the other players to provide more diverse image labels.
Goh et al. compared collaborative and competitive scoring using two versions of the \emph{ESP Game}---one which rewarded players for answer agreement and the other which only rewarded the first player to an agreed answer.
They found no significant differences when looking at player experience metrics and task results, which suggests that rewarding players for competitive behavior had no negative effects on the completion of the task.
Likewise, Siu et al.~\cite{siu2014:gwap} compared collaborative and competitive versions of the game \emph{Cabbage Quest}, finding no significant differences in task completion such as accuracy and rate of task completion, but that players found the competitive version more engaging.
Finally, the previously-mentioned game \emph{Gwario}~\cite{siu2017:gwario} compared collaborative and competitive scoring in its multiplayer condition, finding that players yielded more accurate (but slower) task results in the collaborative version but found the competitive version more challenging.

In addition to examining what players are rewarded \emph{for}, one must also consider what players are rewarded \emph{with}.
Players have different motivations for playing games~\cite{yee2006:motivations} and thus may differ in their preferred kinds of feedback.
Crowdsourcing research has found that players who are intrinsically-motivated to solve tasks are often disengaged by monetary compensation~\cite{mason2010-financial} and that curiosity can be a strong incentive for crowdsourced work~\cite{law2016:curiosity-cat-crowdsourcing}.
This suggests that standard in-game rewards (i.e., points), which may take the place of monetary compensation, may not appeal to all players and that players who are dedicated to the task itself might disengage with games that only provide one kind of reward.
In their analysis of \emph{Foldit}, Cooper et al.~\cite{cooper2010:sci-disc-design} identify a subset of players who are intrinsically motivated to participate in scientific discovery games, which highlights that such intrinsically-motivated players do exist.

So what are the alternatives to points and leaderboards for players who may not find typical \emph{feedback} systems compelling?
Researchers have explored the use of different reward systems.
Goh et al.~\cite{goh2015:reward-systems} compared different versions of a location-based content sharing HCG---one of which rewarded players with points and another which rewarded players with collectible badges---against a control version with no in-game rewards (but which displayed non-gamified progress and statistics).
Overall, few differences in player engagement and task results were found between the versions utilizing points and badges, though both were considered more effective and engaging than the control.
Siu and Riedl~\cite{siu2016:hcg-rewards} investigated multiple reward systems---leaderboards, customizable avatar items, unlockable narrative, and ungamified statistics---using the game \emph{Cafe Flour Sack} in a multivariate experiment with two different player audiences: expert crowdsourced workers and a non-expert student population.
They found that offering players a choice of reward from different systems (e.g., leaderboards, customizable avatar items, and unlockable narrative), as opposed to randomly assigning rewards from these systems, resulted in a better player experience with no differences in task results.
When player audiences were taken into account, expert crowdsourced workers proved to be the most effective, but non-experts could perform nearly as well by being offered the ability to choose which rewards they desired.
Gaston and Cooper~\cite{gaston2017:three-star-foldit} explored the use of three-star reward systems in the context of \emph{Foldit}.
They found that the introduction of three-star reward systems---mechanics awarding players up to three ``stars'' based on their performance of a game level or challenge---resulted in players completing levels in fewer, longer moves.
Additionally players were more likely to replay levels, suggesting that such reward systems could encourage players to interact with the game in desirable ways (e.g., replaying levels to reinforce mastery of concepts in games like \emph{Foldit}, which must teach players strategies necessary to complete the task).
These explorations suggest that different reward systems and their presentation may have effects on the player experience and the completion of the task.
Targeted player audiences may also be affected by how and what kinds of rewards are available, and may behave differently under different conditions.

Finally, we highlight \emph{Project Discovery}~\cite{peplow2016:eve-project-discovery} as a unique example of an HCG that is embedded as a playable experience accessible in a mainstream game: the online game \emph{EVE Online}.
Thematically, the HCG is set in the \emph{EVE Online} universe and players are rewarded with currency that can be spent within the larger game world.
Direct integration of HCGs with existing games (moreover, those designed primarily for entertainment) remains otherwise unexplored, but this example demonstrates a potential way to leverage existing player bases of other games using \emph{feedback} mechanics. 

%% file: methodology.tex
\section{A Methodology for HCG Design}
\label{sec:methodology}
Our mechanics representation provides a breakdown of the different kinds of mechanics in human computation games.
This enables us to identify where we can focus our explorations of the HCG design space, but not \emph{how} we should explore the space in order to build up generalizable design knowledge.

We propose a methodology of controlled A/B design experiments that explore the space of human computation game designs, using formal representations for game elements, tasks, and audiences.
In the context of HCG mechanics, this manifests as between-subjects (alternatively, within-subjects) experiments comparing separate versions of HCGs with different mechanical variations.

These design experiments should (1)~implement a task with a known solution, while (2)~focusing on a single element of a HCG's design.
First, testing with a known solution allows us to evaluate task-related metrics objectively without simultaneously solving a novel problem.
Such known solutions may be the result of pre-solved human computation problems (e.g., image labeling, as discussed in Section \ref{sec:esp-evolution}) or simpler tasks (often those relying on commonsense human knowledge) that are analogous to existing problems.
Second, focusing on one particular element of an HCG's design allows us to understand exactly what kind of impact an element may have on both players and the task with minimal interaction effects.
Our mechanics representation can be used to assist us in understanding where and how the introduction of an element (which may affect any of the \emph{action}, \emph{verification}, and \emph{feedback} mechanics) will affect the HCG game loop.

These experiments should simultaneously evaluate how design decisions meet the needs of players and tasks.
Optimizing only for the player may result in a game with mechanics that do not effectively solve the human computation task even if the game is considered engaging.
Optimizing only for the task may result in a game that players do not find engaging (and thus will not play) even if the game effectively solves the human computation task.
We refer to these two axes of metrics as the \emph{player experience} and the \emph{task completion}.

\emph{Player experience} encompasses metrics such as:
\begin{itemize}
\item \textbf{Engagement}: how players interact with the game or rate their experience with it
\item \textbf{Retention}: how likely are players to continue playing
\item Other subjective measures related to how players interact and perceive the game (e.g., preferences, unstructured self-reported feedback)
\end{itemize}

\emph{Task completion} refers to the task-related metrics such as:
\begin{itemize}
\item \textbf{Quality}: correctness or accuracy of task results
\item \textbf{Volume}: amount of completed tasks
\item \textbf{Diversity}: the variation or breadth of task results
\item \textbf{Rate of Acquisition}: how quickly tasks are completed
\end{itemize}

\noindent The exact metrics to test for often depend on the nature of the human computation task and the HCG's target player audiences.
For example, HCGs for tasks that have a high bar for teaching or training players to solve them effectively may consider metrics such as player retention much more important than HCGs for simpler tasks where maintaining a skilled player base is not a priority.

Moreover, these requirements may change over time.
Task providers may find that their initial \emph{task completion} results may not be sufficient or need additional refinement.
Should they wish to change existing mechanics or introduce new ones, it is imperative to understand how those design changes can ensure the necessary results while still maintaining a positive \emph{player experience}.
Likewise, player audiences (not to mention their preferences) may change, especially if an HCG wishes to broaden its reach to new or different target populations of players.

We note that this methodology is not new, as similar experimental testing approaches have been applied to several instances of HCGs.
Goh et al.~\cite{goh2011:fight-unite} compared a non-gamified control application for image labeling against two versions of the \emph{ESP Game}, one using collaborative scoring mechanisms and the other using competitive scoring mechanisms.
They evaluated both results of the image labeling task and aspects of the application or game experience across these three conditions.
We will further examine this experiment in Section \ref{sec:esp-evolution}.
Similarly, Siu et al.~\cite{siu2014:gwap} conducted an experiment using the game \emph{Cabbage Quest} to test collaborative and competitive scoring mechanisms.
The study utilized a task with a known solution---categorizing everyday objects with purchasing locations---and the controlled experiment compared two in-game scoring mechanisms: one collaborative and one competitive.
Task results were compared to a gold standard to evaluate \emph{task completion} metrics, while player actions and survey responses were logged to evaluate player engagement and experience.
Both Goh et al.'s \emph{ESP Games} and \emph{Cabbage Quest} follow our proposed methodology of taking a problem with a known solution or gold-standard answer set, testing design elements by treating a set of game mechanics as independent variables, and measuring aspects of both the \emph{player experience} and \emph{task completion}.
We note that these experiments benefited from having a gold-standard answer set, but in some cases, a preexisting solution may not be available for new problems.
In such cases, it may be sufficient to evaluate first for \emph{player experience}, followed by later evaluation of \emph{task completion} upon verification of initial task results.


%% file: examples.tex

We now discuss how a controlled methodology of systematically testing game elements in human computation games, combined with our mechanics representation, can be used to study, compare, and explore HCG designs.
We provide two case studies as illustrative examples: (1)~a comparison of image labeling games and studies of their design space and (2)~a discussion of several studies that have utilized (and expanded) this methodology to test multiple aspects of HCGs, including player audiences and novel game elements.

\subsection{Case Study: Comparison and Evolution of Image Labeling Games}
\label{sec:esp-evolution}
\begin{figure*}[tb]
\centering
\includegraphics[width=\textwidth]{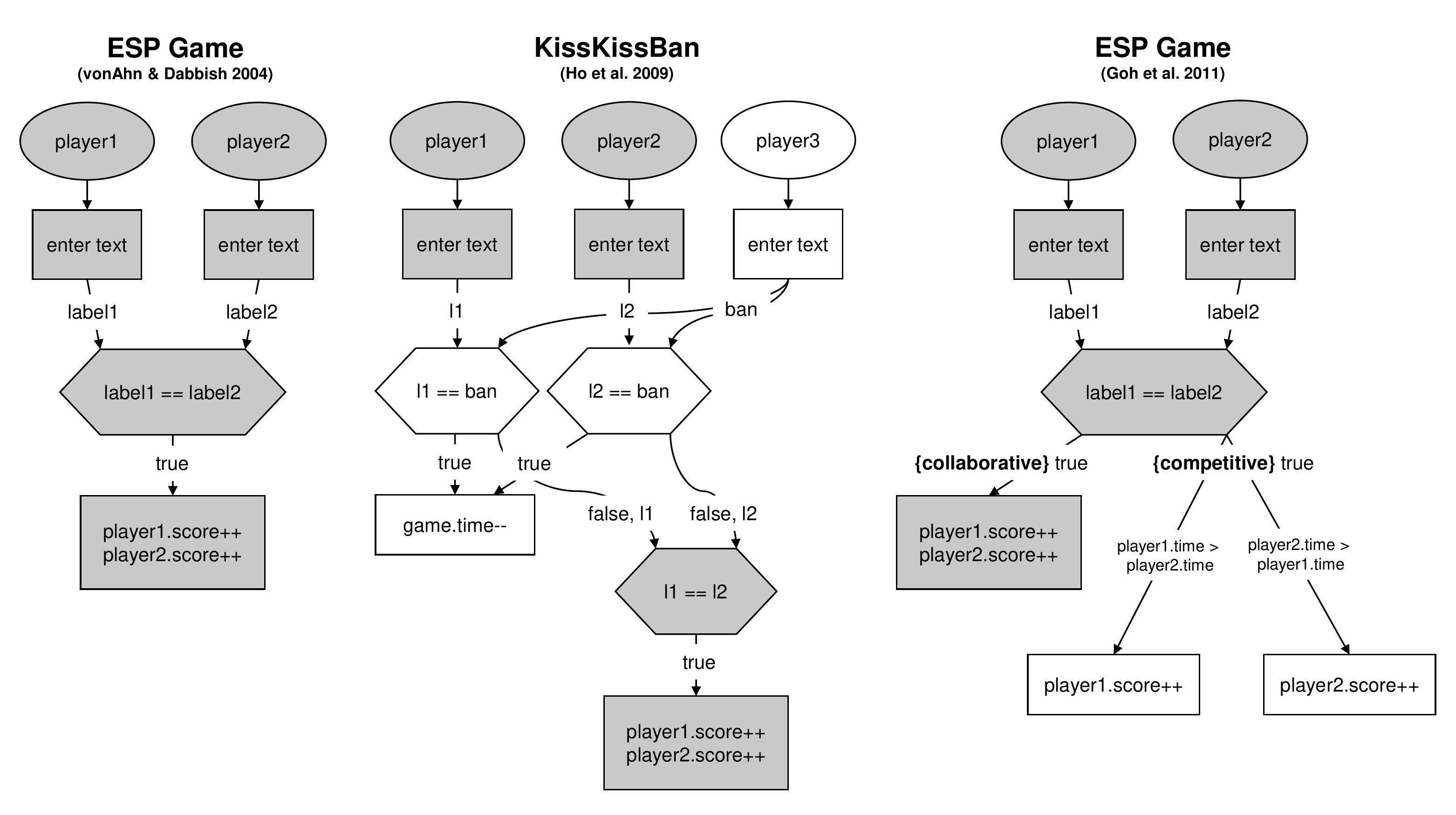}
\caption{Mechanics breakdown of three image labeling HCGs~\cite{vonahn2004:esp, ho2009:kiss-kiss-ban, goh2011:fight-unite}. On the left is the original \emph{ESP Game}, followed by subsequent variations that modified elements of its original design. The mechanics of the original \emph{ESP Game} are colored in gray; novel mechanical variations are colored in white.}
\label{fig:esp-evolution}
\end{figure*}

Image labeling, the task of annotating or classifying images with subject labels or tags, is one of the most iconic and well-studied tasks in human computation games.
Figure \ref{fig:esp-evolution} illustrates our mechanics representation of three HCGs designed to solve the image labeling problem, beginning with the original \emph{ESP Game}~\cite{vonahn2004:esp} followed by \emph{KissKissBan}~\cite{ho2009:kiss-kiss-ban} and Goh et al.'s \emph{ESP Games}~\cite{goh2011:fight-unite}.
Our mechanics representation enables us to identify where changes in game mechanics were made in later games; \autoref{fig:esp-evolution} shows the mechanical structure of the original \emph{ESP Game} highlighted in gray on top of later iterations.
Furthermore, we can visualize how these later games differ, specifically in their explorations of competitive game mechanics.

\emph{KissKissBan}, previously discussed in Section \ref{sec:formalization}, tackled the problem of generating a wider, more diverse set of labels for given images.
While not shown in \autoref{fig:esp-evolution}, the original \emph{ESP Game} eventually adopted the use of ``taboo'' words based on preexisting (repetitive) solutions that players could not input, forcing players to generate a wider range of labels.
In this context, \emph{KissKissBan} provides an alternative mechanical solution to the problem, specifically by introducing an additional player and with it, an additional verification step that did not rely on preexisting labels.
We can see just how these mechanics were injected into the original design of the \emph{ESP Game} in Figure \ref{fig:esp-evolution}.
While the \emph{action} mechanics of entering text input are identical to those of the first two players, the new design incorporates additional verification steps and \emph{feedback} mechanics that can penalize the first two players.

Goh et al., as previously discussed in Section \ref{sec:formalization}, conducted a study comparing collaborative and competitive versions of the \emph{ESP Game} against a non-gamified control application.
Their \emph{ESP Games} are shown Figure \ref{fig:esp-evolution} with the two experimental conditions emphasized in boldfaced braces.
Using our mechanics breakdown, we can identify that the only major mechanical difference between the collaborative and competitive version of their \emph{ESP Games} is in the particular scoring function used to provide feedback behaviors.
Specifically, we can see just how their study isolated and varied one specific \emph{feedback} mechanic: the scoring function.


Understanding how certain mechanics, or collections of mechanics are modified or added on top of existing games can help provide insight to apply them to other HCGs.
Visually, \autoref{fig:esp-evolution} gives us a better understanding of how one might take the changes utilized by \emph{KissKissBan} or Goh et al.'s \emph{ESP Games} and apply them to a structurally-similar \emph{ESP Game} (which vonAhn and Dabbish's templates refer to as ``input-agreement''~\cite{vonahn2008:gwap-design}).

Moreover, this helps to hypothesize just how these collections of mechanics are likely to affect the \emph{player experience} and \emph{task completion}.
Unfortunately, while Goh et al. evaluated both \emph{player experience} and \emph{task completion} metrics in their experiment, \emph{KissKissBan} only evaluated the latter against original \emph{ESP Game}.
This makes it difficult to compare how \emph{KissKissBan's} mechanics might compare to those of Goh et al.'s competitive \emph{ESP Game}, particularly in the context of \emph{player experience}.
Ideally, as more comprehensive design knowledge is built up, HCG developers might be able to treat these collections of mechanics as modular when applying them to new tasks and integrating them into games, assisted by understanding of their potential effects on the \emph{player experience} and \emph{task completion}.

\subsection{Case Study: Exploring Player Audiences and Novel Designs in HCGs}
\label{sec:hcg-studies}

\begin{figure}[tb]
\centering
\includegraphics[width=0.5\textwidth]{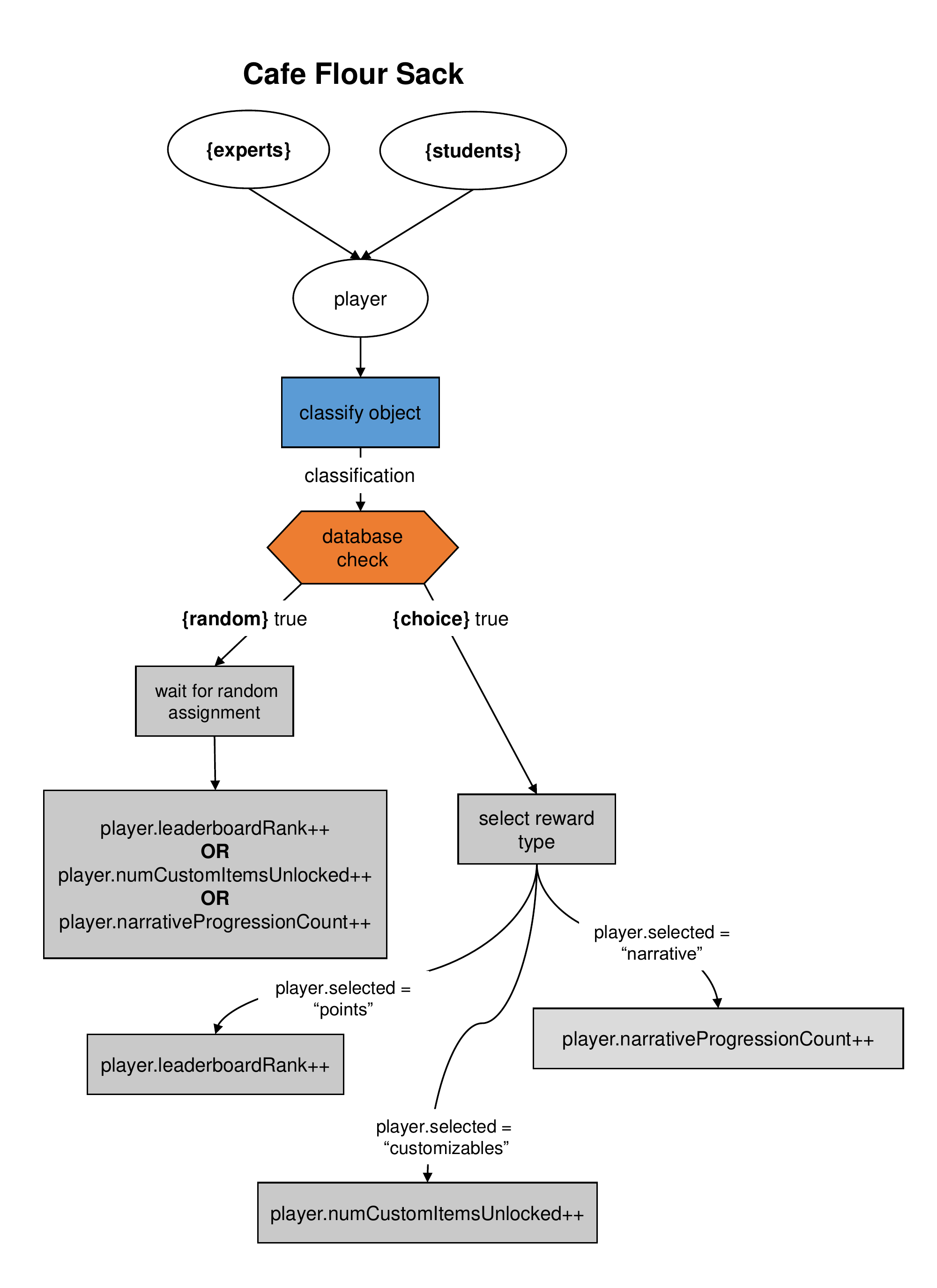}
\caption{Mechanical breakdown of the \emph{Cafe Flour Sack} experiment. Experimental conditions for reward distribution (random versus choice) and player audience (experts versus students) are noted using boldfaced braces.}
\label{fig:cfs}
\end{figure}

As previously described, experiments using HCGs such as the \emph{ESP Game} variations and \emph{Cabbage Quest} have used a controlled methodology as highlighted in Section \ref{sec:methodology} to systematically test aspects of games, in particular \emph{feedback} mechanics.
We now report on two recent experiments that have taken this methodology further by exploring aspects of HCGs such as player audience, considering multivariate conditions (i.e., two independent variables), and challenging longstanding design hypotheses with novel game designs.

\pseudosection{Cafe Flour Sack.}
\emph{Cafe Flour Sack}, previously discussed in Section \ref{sec:feedback-mechanics}, was designed to test aspects of \emph{feedback} mechanics~\cite{siu2016:hcg-rewards}.
The game uses a classification task with a known answer---pairing cooking ingredients with recipes that can be made from them---with drag-and-drop \emph{action} mechanics for players to classify answers.
A visualization of the experiment using our mechanics representation can be seen in Figure \ref{fig:cfs}.

The primary \emph{feedback} mechanic tested was the distribution of rewards: whether players were \emph{randomly} assigned one of three possible reward types or were allowed to \emph{choose} from those reward types.
As shown in Figure \ref{fig:cfs}, the variation occurred only in the \emph{feedback} mechanics.
The only difference between the two versions was a reward selection screen presented to the player: \emph{random}---which showed players a randomly-assigned (highlighted) reward type---and \emph{choice}---which allowed players to choose and click on one of three reward types.
This interface change was kept deliberately minimal, but when looking only at the reward condition, results show that the \emph{choice} condition was perceived more positively by players.
Otherwise, players showed no significant differences in \emph{player experience} or \emph{task completion}.

In addition to testing \emph{random} versus \emph{choice} reward distribution, \emph{Cafe Flour Sack} also tested these conditions across two different player audiences.
One audience consisted of expert crowdsourcing workers recruited on Amazon Mechanical Turk, while the other audience consisted of university students.
Predictably, experts performed overall better than students at \emph{task completion} metrics such as task accuracy/correctness and rate of task completion.
However, when looking at the interactions between the reward conditions, students given a \emph{choice} of rewards performed nearly as well as experts given a \emph{choice} of rewards.
This suggests that under certain conditions such as the appropriate choice of mechanics, non-experts (i.e., students) can perform nearly as well as experts.
We note that \emph{Cafe Flour Sack} uses a simple commonsense knowledge task, but such differences could be more impactful for more complex tasks, although further replication and experimentation is needed to understand how these results hold.

While our proposed methodology focuses on using controlled studies on game \emph{mechanics}, the \emph{Cafe Flour Sack} experiment highlights the need to also consider other aspects of human computation games.
As shown in \autoref{fig:cfs}, our mechanics representation can also be extended to other aspects of HCGs, such as player audience.
The results from \emph{Cafe Flour Sack} suggest that testing conditions such as player audience is just as valuable as testing game mechanics, given the effects on the \emph{player experience} and \emph{task completion} metrics.
We believe that our framework is flexible to such extensions.
For example, while we advocate for fixing the task using this experimental methodology, one could hypothetically explore the effect that different \emph{tasks} have on \emph{player experience} and \emph{task completion} by fixing \emph{mechanics} of an HCG while using similar tasks.

\begin{figure}[tb]
\centering
\includegraphics[width=0.5\textwidth]{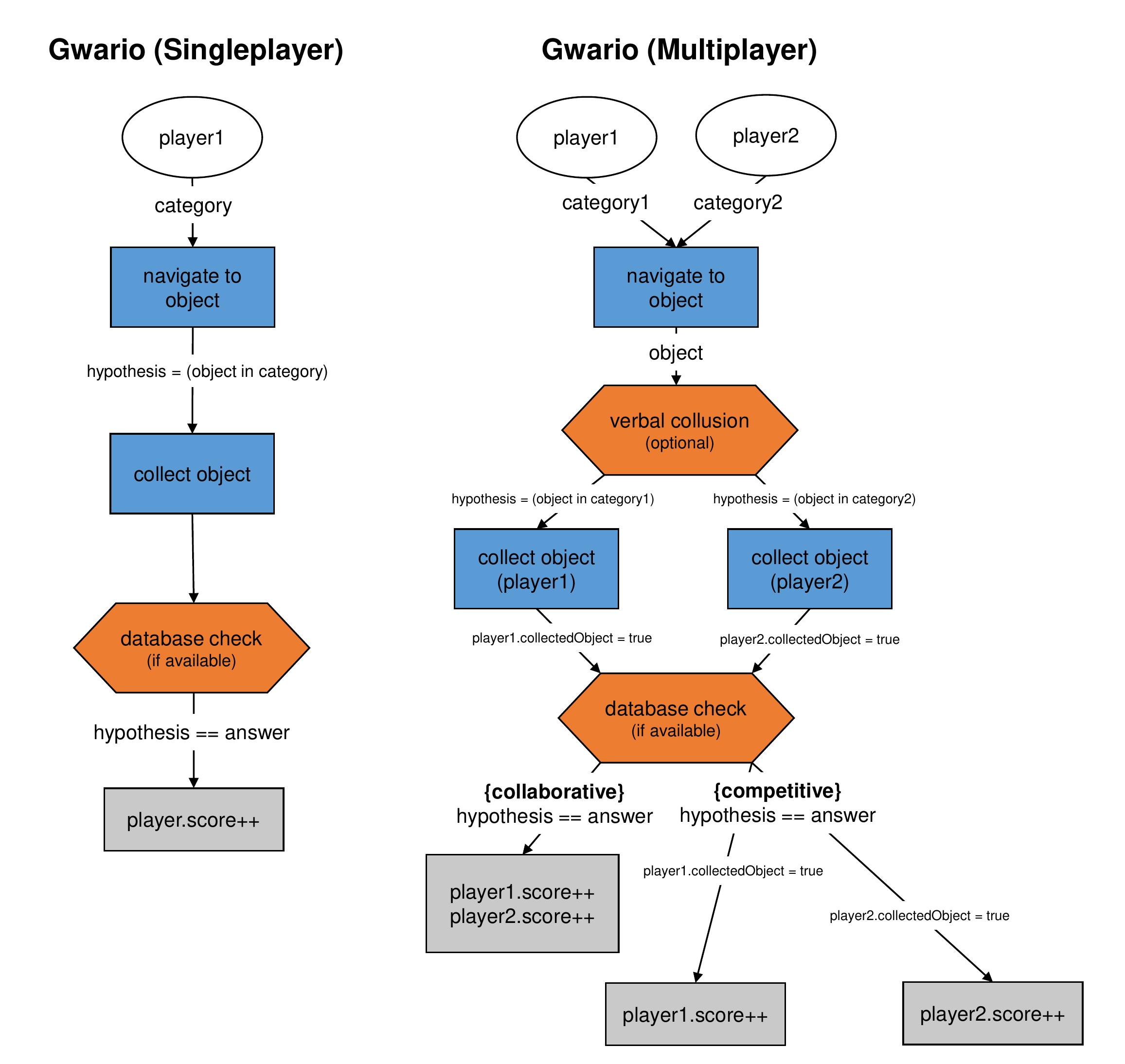}
\caption{Mechanical breakdown of the \emph{Gwario} experiment. Experimental conditions for singleplayer and multiplayer are split for clarity, while multiplayer scoring conditions (collaborative versus competitive) are noted using boldfaced braces.}
\label{fig:gwario}
\end{figure}

\pseudosection{Gwario.}
\emph{Gwario}, previously discussed in Section \ref{sec:formalization}, was designed to test co-located multiplayer game mechanics in the context of an HCG adopting the mechanics of the popular platformer \emph{Super Mario Bros.}~\cite{siu2017:gwario}.
The game utilizes the task with a known solution as that in \emph{Cabbage Quest}~\cite{siu2014:gwap}.
A visualization of the experiment using our mechanics representation can be seen in Figure \ref{fig:gwario}. 

The game tested co-located multiplayer mechanics and consisted of two conditions: a \emph{singleplayer} version and a co-located \emph{multiplayer} version, in which players are seated side-by-side sharing the same game screen.
Figure \ref{fig:gwario} shows that the mechanics of the \emph{singleplayer} version are a deliberately-strict subset of the mechanics in the \emph{multiplayer} version, which adds verbal communication between players as an optional verification step (i.e., players may choose whether or not to talk during gameplay).
Additionally, within its \emph{multiplayer} condition, \emph{Gwario} also compared the experimental conditions described in Goh et al.'s previously-described \emph{ESP Games} and \emph{Cabbage Quest}: collaborative scoring and competitive scoring.
Like these prior experiments, the only variation between the two scoring conditions was the scoring function used to evaluate players.
The only visual difference between the two was the reward screen at the end of the game levels, showing either a joint score or separate scores.


Adoption of mainstream game mechanics still remains a contentious and relatively-unexplored area of human computation game design.
\emph{Gwario's} adaptation of collection and 2D platforming mechanics from \emph{Super Mario Bros.} represents a novel design in the space of HCGs.
This raises the question: how can we safely adopt and test mechanics from successful mainstream games?
When considering a particular genre or game, which mechanics do we adopt?
The \emph{Gwario} experiment demonstrates possible approaches, such adopting a secondary mechanic from \emph{Super Mario Bros.}---coin collection---to the task of classifying objects as part of its \emph{action} mechanics.
Meanwhile, \emph{Super Mario Bros.'} primary navigation mechanics were left untouched to preserve the feel of the original game, but did not affect the completion of the human computation task.
Additionally, \emph{direct communication}---a social element---was treated as a \emph{verification} mechanic.
By breaking HCGs down using our mechanics representation, we can identify and isolate where adopted mainstream game elements could fit into the HCG game loop.
This would allow us to carefully and iteratively test variations while ensuring that design changes to mechanics are isolated to particular parts of the game loop.
Comparison through our proposed methodology then could help to identify potential tradeoffs between the \emph{player experience} and \emph{task completion}.

Additionally, the \emph{Gwario} experiment also highlights the potential for different experimental methods in the context of human computation game design.
\emph{Gwario}'s survey of HCG experts, while limited, provides valuable insights into the design philosophies and considerations of HCG developers.
While our proposed methodology is very amenable to quantitative methods for evaluating games, we believe that HCG design stands to benefit from qualitative and mixed method approaches, such as grounded-theory approaches and/or semi-structured interviews with both players and HCG developers/task providers. 


%% file: conclusions.tex
\section{Conclusions}
In this paper, we describe a framework for designing and studying human computation games.
We believe that generalizable design knowledge for HCGs is necessary, and should consider both players and tasks.
We proposed a formal representation of HCG mechanics into three types: \emph{action}, \emph{verification}, and \emph{feedback}, and illustrate it with examples of its potential use.
This formalization allows us to visualize and study HCG elements, enable formal comparisons of existing HCGs, and facilitate controlled experiments to advance HCG design knowledge.
We highlighted a methodology of running design experiments on known tasks that measure both \emph{player experience} and \emph{task completion} when varying game elements.
We illustrated how to utilize both representation and methodology in two examples: a comparison of image labeling games and a discussion of recent HCG experiments.


Human computation games have demonstrated the potential to solve complex and difficult problems, but must be both engaging experiences for players and effective at solving their tasks.
With the volume and availability of games increasing each day, HCGs must compete for players' time and attention, and thus must remain relevant and consistent with player expectations.
Otherwise, we risk tarnishing their reputation with players, which may ultimately lead to HCGs' premature dismissal as ineffective interfaces for human computation work.
In order to ensure this does not happen, we need to understand how HCGs work, to build better and broader generalizable design knowledge that can adapt to new games, tasks, and audiences, especially when HCG developers do not typically have the training or resources of professional game studios.
Our mechanics representation and our proposed methodology are designed to explore and evaluate HCG mechanics so that it will be easier to design and develop successful, effective HCGs.
In doing so, we hope to work towards a future where HCGs are engaging, effective, ubiquitous, and empowering. 

%% file: paper.bbl

\begin{thebibliography}{00}


\ifx \showCODEN    \undefined \def \showCODEN     #1{\unskip}     \fi
\ifx \showDOI      \undefined \def \showDOI       #1{{\tt DOI:}\penalty0{#1}\ }
  \fi
\ifx \showISBNx    \undefined \def \showISBNx     #1{\unskip}     \fi
\ifx \showISBNxiii \undefined \def \showISBNxiii  #1{\unskip}     \fi
\ifx \showISSN     \undefined \def \showISSN      #1{\unskip}     \fi
\ifx \showLCCN     \undefined \def \showLCCN      #1{\unskip}     \fi
\ifx \shownote     \undefined \def \shownote      #1{#1}          \fi
\ifx \showarticletitle \undefined \def \showarticletitle #1{#1}   \fi
\ifx \showURL      \undefined \def \showURL       #1{#1}          \fi
\providecommand\bibfield[2]{#2}
\providecommand\bibinfo[2]{#2}
\providecommand\natexlab[1]{#1}
\providecommand\showeprint[2][]{arXiv:#2}

\bibitem[\protect\citeauthoryear{Andersen, Liu, Snider, Szeto, Cooper, and
  Popovi\'{c}}{Andersen et~al\mbox{.}}{2011}]%
        {andersen2011:harmful-objectives}
\bibfield{author}{\bibinfo{person}{Erik Andersen}, \bibinfo{person}{Yun-En
  Liu}, \bibinfo{person}{Richard Snider}, \bibinfo{person}{Roy Szeto},
  \bibinfo{person}{Seth Cooper}, {and} \bibinfo{person}{Zoran Popovi\'{c}}.}
  \bibinfo{year}{2011}\natexlab{}.
\newblock \showarticletitle{On the Harmfulness of Secondary Game Objectives}.
  In \bibinfo{booktitle}{{\em Proceedings of the 6th International Conference
  on Foundations of Digital Games}} {\em (\bibinfo{series}{FDG '11})}.
  \bibinfo{publisher}{ACM}, \bibinfo{address}{New York, NY, USA},
  \bibinfo{pages}{30--37}.
\newblock
\showISBNx{978-1-4503-0804-5}
\showDOI{%
\url{http://dx.doi.org/10.1145/2159365.2159370}}


\bibitem[\protect\citeauthoryear{Andersen, O'Rourke, Liu, Snider, Lowdermilk,
  Truong, Cooper, and Popovi\'{c}}{Andersen et~al\mbox{.}}{2012}]%
        {andersen2012:tutorials-impact}
\bibfield{author}{\bibinfo{person}{Erik Andersen}, \bibinfo{person}{Eleanor
  O'Rourke}, \bibinfo{person}{Yun-En Liu}, \bibinfo{person}{Rich Snider},
  \bibinfo{person}{Jeff Lowdermilk}, \bibinfo{person}{David Truong},
  \bibinfo{person}{Seth Cooper}, {and} \bibinfo{person}{Zoran Popovi\'{c}}.}
  \bibinfo{year}{2012}\natexlab{}.
\newblock \showarticletitle{The Impact of Tutorials on Games of Varying
  Complexity}. In \bibinfo{booktitle}{{\em Proceedings of the SIGCHI Conference
  on Human Factors in Computing Systems}} {\em (\bibinfo{series}{CHI '12})}.
  \bibinfo{publisher}{ACM}, \bibinfo{address}{New York, NY, USA},
  \bibinfo{pages}{59--68}.
\newblock
\showISBNx{978-1-4503-1015-4}
\showDOI{%
\url{http://dx.doi.org/10.1145/2207676.2207687}}


\bibitem[\protect\citeauthoryear{Barrington, Turnbull, and
  Lanckriet}{Barrington et~al\mbox{.}}{2012}]%
        {barrington2012:herdit}
\bibfield{author}{\bibinfo{person}{Luke Barrington}, \bibinfo{person}{Douglas
  Turnbull}, {and} \bibinfo{person}{Gert Lanckriet}.}
  \bibinfo{year}{2012}\natexlab{}.
\newblock \showarticletitle{Game-powered machine learning}.
\newblock \bibinfo{journal}{{\em Proceedings of the National Academy of
  Sciences\/}} \bibinfo{volume}{109}, \bibinfo{number}{17}
  (\bibinfo{year}{2012}), \bibinfo{pages}{6411--6416}.
\newblock
\showDOI{%
\url{http://dx.doi.org/10.1073/pnas.1014748109}}
\showeprint{http://www.pnas.org/content/109/17/6411.full.pdf}


\bibitem[\protect\citeauthoryear{Bell, Reeves, Brown, Sherwood, MacMillan,
  Ferguson, and Chalmers}{Bell et~al\mbox{.}}{2009}]%
        {bell2009:eyespy}
\bibfield{author}{\bibinfo{person}{Marek Bell}, \bibinfo{person}{Stuart
  Reeves}, \bibinfo{person}{Barry Brown}, \bibinfo{person}{Scott Sherwood},
  \bibinfo{person}{Donny MacMillan}, \bibinfo{person}{John Ferguson}, {and}
  \bibinfo{person}{Matthew Chalmers}.} \bibinfo{year}{2009}\natexlab{}.
\newblock \showarticletitle{EyeSpy: Supporting Navigation Through Play}. In
  \bibinfo{booktitle}{{\em Proceedings of the SIGCHI Conference on Human
  Factors in Computing Systems}} {\em (\bibinfo{series}{CHI '09})}.
  \bibinfo{publisher}{ACM}, \bibinfo{address}{New York, NY, USA},
  \bibinfo{pages}{123--132}.
\newblock
\showISBNx{978-1-60558-246-7}
\showDOI{%
\url{http://dx.doi.org/10.1145/1518701.1518723}}


\bibitem[\protect\citeauthoryear{Cooper, Khatib, Makedon, Lu, Barbero, Baker,
  Fogarty, Popovi\'{c}, and players}{Cooper et~al\mbox{.}}{2011}]%
        {cooper2011:foldit-recipes}
\bibfield{author}{\bibinfo{person}{Seth Cooper}, \bibinfo{person}{Firas
  Khatib}, \bibinfo{person}{Ilya Makedon}, \bibinfo{person}{Hao Lu},
  \bibinfo{person}{Janos Barbero}, \bibinfo{person}{David Baker},
  \bibinfo{person}{James Fogarty}, \bibinfo{person}{Zoran Popovi\'{c}}, {and}
  \bibinfo{person}{Foldit players}.} \bibinfo{year}{2011}\natexlab{}.
\newblock \showarticletitle{Analysis of Social Gameplay Macros in the Foldit
  Cookbook}. In \bibinfo{booktitle}{{\em Proceedings of the 6th International
  Conference on Foundations of Digital Games}} {\em (\bibinfo{series}{FDG
  '11})}. \bibinfo{publisher}{ACM}, \bibinfo{address}{New York, NY, USA},
  \bibinfo{pages}{9--14}.
\newblock
\showISBNx{978-1-4503-0804-5}


\bibitem[\protect\citeauthoryear{Cooper, Khatib, Treuille, Barbero, Lee,
  Beenen, Leaver-Fay, Baker, Popovi{\'c}, et~al\mbox{.}}{Cooper
  et~al\mbox{.}}{2010a}]%
        {cooper2010:foldit}
\bibfield{author}{\bibinfo{person}{Seth Cooper}, \bibinfo{person}{Firas
  Khatib}, \bibinfo{person}{Adrien Treuille}, \bibinfo{person}{Janos Barbero},
  \bibinfo{person}{Jeehyung Lee}, \bibinfo{person}{Michael Beenen},
  \bibinfo{person}{Andrew Leaver-Fay}, \bibinfo{person}{David Baker},
  \bibinfo{person}{Zoran Popovi{\'c}}, {and} \bibinfo{person}{others}.}
  \bibinfo{year}{2010}\natexlab{a}.
\newblock \showarticletitle{Predicting protein structures with a multiplayer
  online game}.
\newblock \bibinfo{journal}{{\em Nature\/}} \bibinfo{volume}{466},
  \bibinfo{number}{7307} (\bibinfo{year}{2010}), \bibinfo{pages}{756--760}.
\newblock


\bibitem[\protect\citeauthoryear{Cooper, Treuille, Barbero, Leaver-Fay, Tuite,
  Khatib, Snyder, Beenen, Salesin, Baker, Popovi\'c, and Players}{Cooper
  et~al\mbox{.}}{2010b}]%
        {cooper2010:sci-disc-design}
\bibfield{author}{\bibinfo{person}{Seth Cooper}, \bibinfo{person}{Adrien
  Treuille}, \bibinfo{person}{Janos Barbero}, \bibinfo{person}{Andrew
  Leaver-Fay}, \bibinfo{person}{Kathleen Tuite}, \bibinfo{person}{Firas
  Khatib}, \bibinfo{person}{Alex~Cho Snyder}, \bibinfo{person}{Michael Beenen},
  \bibinfo{person}{David Salesin}, \bibinfo{person}{David Baker},
  \bibinfo{person}{Zoran Popovi\'c}, {and} \bibinfo{person}{Foldit Players}.}
  \bibinfo{year}{2010}\natexlab{b}.
\newblock \showarticletitle{The Challenge of Designing Scientific Discovery
  Games}. In \bibinfo{booktitle}{{\em 5th International Conference on the
  Foundations of Digital Games}}.
\newblock


\bibitem[\protect\citeauthoryear{Dietl, Dietzel, Ernst, Mote, Walker, Cooper,
  Pavlik, and Popovi\'{c}}{Dietl et~al\mbox{.}}{2012}]%
        {dietl2012:verigames}
\bibfield{author}{\bibinfo{person}{Werner Dietl}, \bibinfo{person}{Stephanie
  Dietzel}, \bibinfo{person}{Michael~D. Ernst}, \bibinfo{person}{Nathaniel
  Mote}, \bibinfo{person}{Brian Walker}, \bibinfo{person}{Seth Cooper},
  \bibinfo{person}{Timothy Pavlik}, {and} \bibinfo{person}{Zoran Popovi\'{c}}.}
  \bibinfo{year}{2012}\natexlab{}.
\newblock \showarticletitle{Verification Games: Making Verification Fun}. In
  \bibinfo{booktitle}{{\em 14th Workshop on Formal Techniques for Java-like
  Programs}}.
\newblock


\bibitem[\protect\citeauthoryear{Doroudi, Kamar, Brunskill, and
  Horvitz}{Doroudi et~al\mbox{.}}{2016}]%
        {doroudi2016:towards-learning-science-for-crowdsourcing}
\bibfield{author}{\bibinfo{person}{Shayan Doroudi}, \bibinfo{person}{Ece
  Kamar}, \bibinfo{person}{Emma Brunskill}, {and} \bibinfo{person}{Eric
  Horvitz}.} \bibinfo{year}{2016}\natexlab{}.
\newblock \showarticletitle{Toward a Learning Science for Complex Crowdsourcing
  Tasks}. In \bibinfo{booktitle}{{\em Proceedings of the 2016 CHI Conference on
  Human Factors in Computing Systems}} {\em (\bibinfo{series}{CHI '16})}.
  \bibinfo{publisher}{ACM}, \bibinfo{address}{New York, NY, USA},
  \bibinfo{pages}{2623--2634}.
\newblock
\showISBNx{978-1-4503-3362-7}
\showDOI{%
\url{http://dx.doi.org/10.1145/2858036.2858268}}


\bibitem[\protect\citeauthoryear{Frazier and Riedl}{Frazier and Riedl}{2014}]%
        {frazier2014:proactive-sensing}
\bibfield{author}{\bibinfo{person}{Spencer Frazier} {and} \bibinfo{person}{Mark
  Riedl}.} \bibinfo{year}{2014}\natexlab{}.
\newblock \showarticletitle{Persistent and Pervasive Real-World Sensing Using
  Games}. In \bibinfo{booktitle}{{\em Second AAAI Conference on Human
  Computation and Crowdsourcing}}.
\newblock


\bibitem[\protect\citeauthoryear{Gaston and Cooper}{Gaston and Cooper}{2017}]%
        {gaston2017:three-star-foldit}
\bibfield{author}{\bibinfo{person}{Jacqueline Gaston} {and}
  \bibinfo{person}{Seth Cooper}.} \bibinfo{year}{2017}\natexlab{}.
\newblock \showarticletitle{To Three or Not to Three: Improving Human
  Computation Game Onboarding with a Three-Star System}. In
  \bibinfo{booktitle}{{\em Proceedings of the 2017 CHI Conference on Human
  Factors in Computing Systems}} {\em (\bibinfo{series}{CHI '17})}.
  \bibinfo{publisher}{ACM}, \bibinfo{address}{New York, NY, USA},
  \bibinfo{pages}{5034--5039}.
\newblock
\showISBNx{978-1-4503-4655-9}
\showDOI{%
\url{http://dx.doi.org/10.1145/3025453.3025997}}


\bibitem[\protect\citeauthoryear{Goh, Ang, Lee, and Chua}{Goh
  et~al\mbox{.}}{2011}]%
        {goh2011:fight-unite}
\bibfield{author}{\bibinfo{person}{Dion Hoe-Lian Goh},
  \bibinfo{person}{Rebecca~P. Ang}, \bibinfo{person}{Chei~Sian Lee}, {and}
  \bibinfo{person}{Alton Y.~K. Chua}.} \bibinfo{year}{2011}\natexlab{}.
\newblock \showarticletitle{Fight or Unite: Investigating Game Genres for Image
  Tagging}.
\newblock \bibinfo{journal}{{\em J. Am. Soc. Inf. Sci. Technol.\/}}
  \bibinfo{volume}{62}, \bibinfo{number}{7} (\bibinfo{date}{July}
  \bibinfo{year}{2011}), \bibinfo{pages}{1311--1324}.
\newblock
\showISSN{1532-2882}


\bibitem[\protect\citeauthoryear{Goh, Pe-Than, and Lee}{Goh
  et~al\mbox{.}}{2015}]%
        {goh2015:reward-systems}
\bibfield{author}{\bibinfo{person}{Dion Hoe-Lian Goh},
  \bibinfo{person}{Ei~Pa~Pa Pe-Than}, {and} \bibinfo{person}{Chei~Sian Lee}.}
  \bibinfo{year}{2015}\natexlab{}.
\newblock \bibinfo{booktitle}{{\em An Investigation of Reward Systems in Human
  Computation Games}}.
\newblock \bibinfo{publisher}{Springer International Publishing},
  \bibinfo{pages}{596--607}.
\newblock
\showISBNx{978-3-319-20916-6}


\bibitem[\protect\citeauthoryear{Ho, Chang, Lee, Hsu, and Chen}{Ho
  et~al\mbox{.}}{2010}]%
        {ho2009:kiss-kiss-ban}
\bibfield{author}{\bibinfo{person}{Chien-Ju Ho}, \bibinfo{person}{Tao-Hsuan
  Chang}, \bibinfo{person}{Jong-Chuan Lee}, \bibinfo{person}{Jane Yung-jen
  Hsu}, {and} \bibinfo{person}{Kuan-Ta Chen}.} \bibinfo{year}{2010}\natexlab{}.
\newblock \showarticletitle{KissKissBan: A Competitive Human Computation Game
  for Image Annotation}. \bibinfo{journal}{{\em SIGKDD Explor. Newsl.\/}}
  \bibinfo{volume}{12}, \bibinfo{number}{1} (\bibinfo{date}{Nov.}
  \bibinfo{year}{2010}), \bibinfo{pages}{21--24}.
\newblock
\showISSN{1931-0145}
\showDOI{%
\url{http://dx.doi.org/10.1145/1882471.1882475}}


\bibitem[\protect\citeauthoryear{Isbister and Schaffer}{Isbister and
  Schaffer}{2008}]%
        {isbister2008:game-usability}
\bibfield{author}{\bibinfo{person}{Katherine Isbister} {and}
  \bibinfo{person}{Noah Schaffer}.} \bibinfo{year}{2008}\natexlab{}.
\newblock \bibinfo{booktitle}{{\em Game Usability: Advancing the Player
  Experience}}.
\newblock \bibinfo{publisher}{Morgan Kaufmann}.
\newblock


\bibitem[\protect\citeauthoryear{Jamieson, Grace, and Hall}{Jamieson
  et~al\mbox{.}}{2012}]%
        {jamieson2012:hcomp-games}
\bibfield{author}{\bibinfo{person}{Peter Jamieson}, \bibinfo{person}{Lindsay
  Grace}, {and} \bibinfo{person}{Jack Hall}.} \bibinfo{year}{2012}\natexlab{}.
\newblock \showarticletitle{Research Directions for Pushing Harnessing Human
  Computation to Mainstream Video Games}. In \bibinfo{booktitle}{{\em
  Meaningful Play}}.
\newblock


\bibitem[\protect\citeauthoryear{Kawrykow, Roumanis, Kam, Kwak, Leung, Wu,
  Zarour, Sarmenta, Blanchette, Waldisp{\"u}hl, and Players}{Kawrykow
  et~al\mbox{.}}{2012}]%
        {kawrykow2012:phylo}
\bibfield{author}{\bibinfo{person}{Alexander Kawrykow}, \bibinfo{person}{Gary
  Roumanis}, \bibinfo{person}{Alfred Kam}, \bibinfo{person}{Daniel Kwak},
  \bibinfo{person}{Clarence Leung}, \bibinfo{person}{Chu Wu},
  \bibinfo{person}{Eleyine Zarour}, \bibinfo{person}{Luis Sarmenta},
  \bibinfo{person}{Mathieu Blanchette}, \bibinfo{person}{J{\'e}r{\^o}me
  Waldisp{\"u}hl}, {and} \bibinfo{person}{Phylo Players}.}
  \bibinfo{year}{2012}\natexlab{}.
\newblock \showarticletitle{Phylo: a citizen science approach for improving
  multiple sequence alignment}.
\newblock \bibinfo{journal}{{\em PLoS One\/}} \bibinfo{volume}{7},
  \bibinfo{number}{3} (\bibinfo{year}{2012}), \bibinfo{pages}{e31362}.
\newblock


\bibitem[\protect\citeauthoryear{Krause and Smeddinck}{Krause and
  Smeddinck}{2011}]%
        {krause2011:hcg-survey}
\bibfield{author}{\bibinfo{person}{Markus Krause} {and} \bibinfo{person}{Jan
  Smeddinck}.} \bibinfo{year}{2011}\natexlab{}.
\newblock \showarticletitle{Human computation games: A survey}. In
  \bibinfo{booktitle}{{\em 2011 19th European Signal Processing Conference}}.
  IEEE, \bibinfo{pages}{754--758}.
\newblock
\showISSN{2076-1465}


\bibitem[\protect\citeauthoryear{Krause, Takhtamysheva, Wittstock, and
  Malaka}{Krause et~al\mbox{.}}{2010}]%
        {krause2010:ontogalaxy}
\bibfield{author}{\bibinfo{person}{Markus Krause}, \bibinfo{person}{Aneta
  Takhtamysheva}, \bibinfo{person}{Marion Wittstock}, {and}
  \bibinfo{person}{Rainer Malaka}.} \bibinfo{year}{2010}\natexlab{}.
\newblock \showarticletitle{Frontiers of a Paradigm: Exploring Human
  Computation with Digital Games}. In \bibinfo{booktitle}{{\em Proceedings of
  the ACM SIGKDD Workshop on Human Computation}} {\em (\bibinfo{series}{HCOMP
  '10})}. \bibinfo{publisher}{ACM}, \bibinfo{address}{New York, NY, USA},
  \bibinfo{pages}{22--25}.
\newblock
\showISBNx{978-1-4503-0222-7}
\showDOI{%
\url{http://dx.doi.org/10.1145/1837885.1837893}}


\bibitem[\protect\citeauthoryear{Kuo, Lee, Chiang, Wang, Shen, Chan, and
  Hsu}{Kuo et~al\mbox{.}}{2009}]%
        {kuo2009:virtualpetgame}
\bibfield{author}{\bibinfo{person}{Yen-ling Kuo}, \bibinfo{person}{Jong-Chuan
  Lee}, \bibinfo{person}{Kai-yang Chiang}, \bibinfo{person}{Rex Wang},
  \bibinfo{person}{Edward Shen}, \bibinfo{person}{Cheng-wei Chan}, {and}
  \bibinfo{person}{Jane Yung-jen Hsu}.} \bibinfo{year}{2009}\natexlab{}.
\newblock \showarticletitle{Community-based Game Design: Experiments on Social
  Games for Commonsense Data Collection}. In \bibinfo{booktitle}{{\em
  Proceedings of the ACM SIGKDD Workshop on Human Computation}} {\em
  (\bibinfo{series}{HCOMP '09})}. \bibinfo{publisher}{ACM},
  \bibinfo{address}{New York, NY, USA}, \bibinfo{pages}{15--22}.
\newblock
\showISBNx{978-1-60558-672-4}
\showDOI{%
\url{http://dx.doi.org/10.1145/1600150.1600154}}


\bibitem[\protect\citeauthoryear{Lankes, Mirlacher, Wagner, and
  Hochleitner}{Lankes et~al\mbox{.}}{2014}]%
        {lankes2014:player-gaze}
\bibfield{author}{\bibinfo{person}{Michael Lankes}, \bibinfo{person}{Thomas
  Mirlacher}, \bibinfo{person}{Stefan Wagner}, {and} \bibinfo{person}{Wolfgang
  Hochleitner}.} \bibinfo{year}{2014}\natexlab{}.
\newblock \showarticletitle{Whom Are You Looking for?: The Effects of Different
  Player Representation Relations on the Presence in Gaze-based Games}. In
  \bibinfo{booktitle}{{\em Proceedings of the First ACM SIGCHI Annual Symposium
  on Computer-human Interaction in Play}} {\em (\bibinfo{series}{CHI PLAY
  '14})}. \bibinfo{publisher}{ACM}, \bibinfo{address}{New York, NY, USA},
  \bibinfo{pages}{171--179}.
\newblock
\showISBNx{978-1-4503-3014-5}
\showDOI{%
\url{http://dx.doi.org/10.1145/2658537.2658698}}


\bibitem[\protect\citeauthoryear{Law and von Ahn}{Law and von Ahn}{2009}]%
        {law2009:tagatune}
\bibfield{author}{\bibinfo{person}{Edith Law} {and} \bibinfo{person}{Luis von
  Ahn}.} \bibinfo{year}{2009}\natexlab{}.
\newblock \showarticletitle{Input-agreement: A New Mechanism for Collecting
  Data Using Human Computation Games}. In \bibinfo{booktitle}{{\em Proceedings
  of the SIGCHI Conference on Human Factors in Computing Systems}} {\em
  (\bibinfo{series}{CHI '09})}. \bibinfo{publisher}{ACM}, \bibinfo{address}{New
  York, NY, USA}, \bibinfo{pages}{1197--1206}.
\newblock
\showISBNx{978-1-60558-246-7}
\showDOI{%
\url{http://dx.doi.org/10.1145/1518701.1518881}}


\bibitem[\protect\citeauthoryear{Law and von Ahn}{Law and von Ahn}{2011}]%
        {law2011:hcomp-book}
\bibfield{author}{\bibinfo{person}{Edith Law} {and} \bibinfo{person}{Luis von
  Ahn}.} \bibinfo{year}{2011}\natexlab{}.
\newblock \showarticletitle{Human computation}.
\newblock \bibinfo{journal}{{\em Synthesis Lectures on Artificial Intelligence
  and Machine Learning\/}} \bibinfo{volume}{5}, \bibinfo{number}{3}
  (\bibinfo{year}{2011}), \bibinfo{pages}{1--121}.
\newblock


\bibitem[\protect\citeauthoryear{Law, Yin, Goh, Chen, Terry, and Gajos}{Law
  et~al\mbox{.}}{2016}]%
        {law2016:curiosity-cat-crowdsourcing}
\bibfield{author}{\bibinfo{person}{Edith Law}, \bibinfo{person}{Ming Yin},
  \bibinfo{person}{Joslin Goh}, \bibinfo{person}{Kevin Chen},
  \bibinfo{person}{Michael~A. Terry}, {and} \bibinfo{person}{Krzysztof~Z.
  Gajos}.} \bibinfo{year}{2016}\natexlab{}.
\newblock \showarticletitle{Curiosity Killed the Cat, but Makes Crowdwork
  Better}. In \bibinfo{booktitle}{{\em Proceedings of the 2016 CHI Conference
  on Human Factors in Computing Systems}} {\em (\bibinfo{series}{CHI '16})}.
  \bibinfo{publisher}{ACM}, \bibinfo{address}{New York, NY, USA},
  \bibinfo{pages}{4098--4110}.
\newblock
\showISBNx{978-1-4503-3362-7}
\showDOI{%
\url{http://dx.doi.org/10.1145/2858036.2858144}}


\bibitem[\protect\citeauthoryear{Lee, Kladwang, Lee, Cantu, Azizyan, Kim,
  Limpaecher, Yoon, Treuille, Das, and Participants}{Lee et~al\mbox{.}}{2014}]%
        {lee2014:eterna}
\bibfield{author}{\bibinfo{person}{Jeehyung Lee}, \bibinfo{person}{Wipapat
  Kladwang}, \bibinfo{person}{Minjae Lee}, \bibinfo{person}{Daniel Cantu},
  \bibinfo{person}{Martin Azizyan}, \bibinfo{person}{Hanjoo Kim},
  \bibinfo{person}{Alex Limpaecher}, \bibinfo{person}{Sungroh Yoon},
  \bibinfo{person}{Adrien Treuille}, \bibinfo{person}{Rhiju Das}, {and}
  \bibinfo{person}{EteRNA Participants}.} \bibinfo{year}{2014}\natexlab{}.
\newblock \showarticletitle{{RNA} design rules from a massive open laboratory}.
\newblock \bibinfo{journal}{{\em Proceedings of the National Academy of
  Sciences\/}} \bibinfo{volume}{111}, \bibinfo{number}{6}
  (\bibinfo{year}{2014}), \bibinfo{pages}{2122--2127}.
\newblock


\bibitem[\protect\citeauthoryear{Lintott, Schawinski, Slosar, Land, Bamford,
  Thomas, Raddick, Nichol, Szalay, Andreescu, Murray, and Vandenberg}{Lintott
  et~al\mbox{.}}{2008}]%
        {lintott2008:galaxyzoo}
\bibfield{author}{\bibinfo{person}{Chris~J Lintott}, \bibinfo{person}{Kevin
  Schawinski}, \bibinfo{person}{An{\v{z}}e Slosar}, \bibinfo{person}{Kate
  Land}, \bibinfo{person}{Steven Bamford}, \bibinfo{person}{Daniel Thomas},
  \bibinfo{person}{M~Jordan Raddick}, \bibinfo{person}{Robert~C Nichol},
  \bibinfo{person}{Alex Szalay}, \bibinfo{person}{Dan Andreescu},
  \bibinfo{person}{Phil Murray}, {and} \bibinfo{person}{Jan Vandenberg}.}
  \bibinfo{year}{2008}\natexlab{}.
\newblock \showarticletitle{{Galaxy Zoo}: morphologies derived from visual
  inspection of galaxies from the Sloan Digital Sky Survey}.
\newblock \bibinfo{journal}{{\em Monthly Notices of the Royal Astronomical
  Society\/}} \bibinfo{volume}{389}, \bibinfo{number}{3}
  (\bibinfo{year}{2008}), \bibinfo{pages}{1179--1189}.
\newblock


\bibitem[\protect\citeauthoryear{Liu, Mandel, Brunskill, and Popovi\'{c}}{Liu
  et~al\mbox{.}}{2014}]%
        {liu2014:knowl-vs-learning-bandit}
\bibfield{author}{\bibinfo{person}{Yun-En Liu}, \bibinfo{person}{Travis
  Mandel}, \bibinfo{person}{Emma Brunskill}, {and} \bibinfo{person}{Zoran
  Popovi\'{c}}.} \bibinfo{year}{2014}\natexlab{}.
\newblock \showarticletitle{Trading Off Scientific Knowledge and User Learning
  with Multi-Armed Bandits}. In \bibinfo{booktitle}{{\em Educational Data
  Mining}}.
\newblock


\bibitem[\protect\citeauthoryear{Logas, Whitehead, Mateas, Vallejos, Scott,
  Shapiro, Murray, Compton, Osborn, Salvatore, Lin, Sanchez, Shavlovsky,
  Cetina, Clementi, and Lewis}{Logas et~al\mbox{.}}{2014}]%
        {logas2014:xylem}
\bibfield{author}{\bibinfo{person}{Heather Logas}, \bibinfo{person}{Jim
  Whitehead}, \bibinfo{person}{Michael Mateas}, \bibinfo{person}{Richard
  Vallejos}, \bibinfo{person}{Lauren Scott}, \bibinfo{person}{Dan Shapiro},
  \bibinfo{person}{John Murray}, \bibinfo{person}{Kate Compton},
  \bibinfo{person}{Joseph Osborn}, \bibinfo{person}{Orlando Salvatore},
  \bibinfo{person}{Zhongpeng Lin}, \bibinfo{person}{Huascar Sanchez},
  \bibinfo{person}{Michael Shavlovsky}, \bibinfo{person}{Daniel Cetina},
  \bibinfo{person}{Shayne Clementi}, {and} \bibinfo{person}{Chris Lewis}.}
  \bibinfo{year}{2014}\natexlab{}.
\newblock \showarticletitle{Software Verification Games: Designing {X}ylem, The
  Code of Plants}. In \bibinfo{booktitle}{{\em Proceedings of the 9th
  International Conference on the Foundations of Digital Games}}.
\newblock


\bibitem[\protect\citeauthoryear{Lomas, Patel, Forlizzi, and Koedinger}{Lomas
  et~al\mbox{.}}{2013}]%
        {lomas2013:opt-edugame-challenge}
\bibfield{author}{\bibinfo{person}{Derek Lomas}, \bibinfo{person}{Kishan
  Patel}, \bibinfo{person}{Jodi~L. Forlizzi}, {and} \bibinfo{person}{Kenneth~R.
  Koedinger}.} \bibinfo{year}{2013}\natexlab{}.
\newblock \showarticletitle{Optimizing Challenge in an Educational Game Using
  Large-scale Design Experiments}. In \bibinfo{booktitle}{{\em Proceedings of
  the SIGCHI Conference on Human Factors in Computing Systems}} {\em
  (\bibinfo{series}{CHI '13})}. \bibinfo{publisher}{ACM}, \bibinfo{address}{New
  York, NY, USA}, \bibinfo{pages}{89--98}.
\newblock
\showISBNx{978-1-4503-1899-0}
\showDOI{%
\url{http://dx.doi.org/10.1145/2470654.2470668}}


\bibitem[\protect\citeauthoryear{Mason and Watts}{Mason and Watts}{2010}]%
        {mason2010-financial}
\bibfield{author}{\bibinfo{person}{Winter Mason} {and}
  \bibinfo{person}{Duncan~J. Watts}.} \bibinfo{year}{2010}\natexlab{}.
\newblock \showarticletitle{Financial Incentives and the "Performance of
  Crowds"}.
\newblock \bibinfo{journal}{{\em SIGKDD Explor. Newsl.\/}}
  \bibinfo{volume}{11}, \bibinfo{number}{2} (\bibinfo{date}{May}
  \bibinfo{year}{2010}), \bibinfo{pages}{100--108}.
\newblock
\showISSN{1931-0145}
\showDOI{%
\url{http://dx.doi.org/10.1145/1809400.1809422}}


\bibitem[\protect\citeauthoryear{McEwan, Blackler, Johnson, and Wyeth}{McEwan
  et~al\mbox{.}}{2014}]%
        {mcewan2014:natural-mapping}
\bibfield{author}{\bibinfo{person}{Mitchell~W. McEwan},
  \bibinfo{person}{Alethea~L. Blackler}, \bibinfo{person}{Daniel~M. Johnson},
  {and} \bibinfo{person}{Peta~A. Wyeth}.} \bibinfo{year}{2014}\natexlab{}.
\newblock \showarticletitle{Natural Mapping and Intuitive Interaction in
  Videogames}. In \bibinfo{booktitle}{{\em Proceedings of the First ACM SIGCHI
  Annual Symposium on Computer-human Interaction in Play}} {\em
  (\bibinfo{series}{CHI PLAY '14})}. \bibinfo{publisher}{ACM},
  \bibinfo{address}{New York, NY, USA}, \bibinfo{pages}{191--200}.
\newblock
\showISBNx{978-1-4503-3014-5}
\showDOI{%
\url{http://dx.doi.org/10.1145/2658537.2658541}}


\bibitem[\protect\citeauthoryear{Mehta, Crawford, Luo, Parde, Patel, Rodgers,
  Sistla, Yadav, and Reisner}{Mehta et~al\mbox{.}}{2013}]%
        {mehta2013:untangled-mapping-game}
\bibfield{author}{\bibinfo{person}{Gayatri Mehta}, \bibinfo{person}{Carson
  Crawford}, \bibinfo{person}{Xiaozhong Luo}, \bibinfo{person}{Natalie Parde},
  \bibinfo{person}{Krunalkumar Patel}, \bibinfo{person}{Brandon Rodgers},
  \bibinfo{person}{Anil~Kumar Sistla}, \bibinfo{person}{Anil Yadav}, {and}
  \bibinfo{person}{Marc Reisner}.} \bibinfo{year}{2013}\natexlab{}.
\newblock \showarticletitle{UNTANGLED: A Game Environment for Discovery of
  Creative Mapping Strategies}.
\newblock \bibinfo{journal}{{\em ACM Trans. Reconfigurable Technol. Syst.\/}}
  \bibinfo{volume}{6}, \bibinfo{number}{3}, Article \bibinfo{articleno}{13}
  (\bibinfo{date}{Oct.} \bibinfo{year}{2013}), \bibinfo{numpages}{26}~pages.
\newblock
\showISSN{1936-7406}
\showDOI{%
\url{http://dx.doi.org/10.1145/2517325}}


\bibitem[\protect\citeauthoryear{Pe-Than, Goh, and Lee}{Pe-Than
  et~al\mbox{.}}{2013}]%
        {pe-than2013:hcg-typology}
\bibfield{author}{\bibinfo{person}{Ei~Pa~Pa Pe-Than}, \bibinfo{person}{Dion
  Hoe-Lian Goh}, {and} \bibinfo{person}{Chei~Sian Lee}.}
  \bibinfo{year}{2013}\natexlab{}.
\newblock \showarticletitle{A typology of human computation games: an analysis
  and a review of current games}.
\newblock \bibinfo{journal}{{\em Behaviour \& Information Technology\/}}
  (\bibinfo{year}{2013}).
\newblock


\bibitem[\protect\citeauthoryear{Peplow}{Peplow}{2016}]%
        {peplow2016:eve-project-discovery}
\bibfield{author}{\bibinfo{person}{Mark Peplow}.}
  \bibinfo{year}{2016}\natexlab{}.
\newblock \showarticletitle{Citizen science lures gamers into Sweden's Human
  Protein Atlas}.
\newblock \bibinfo{journal}{{\em Nature Biotechnology\/}} \bibinfo{volume}{34},
  \bibinfo{number}{5} (\bibinfo{year}{2016}), \bibinfo{pages}{452--452}.
\newblock
\showURL{%
\url{http://dx.doi.org/10.1038/nbt0516-452c}}


\bibitem[\protect\citeauthoryear{Seif El-Nasr, Drachen, and Canossa}{Seif
  El-Nasr et~al\mbox{.}}{2013}]%
        {seifel-nasr2013:game-analytics-book}
\bibfield{editor}{\bibinfo{person}{Magy Seif El-Nasr}, \bibinfo{person}{Anders
  Drachen}, {and} \bibinfo{person}{Alessandro Canossa}} (Eds.).
  \bibinfo{year}{2013}\natexlab{}.
\newblock \bibinfo{booktitle}{{\em Game Analytics}}.
\newblock \bibinfo{publisher}{Springer London}.
\newblock


\bibitem[\protect\citeauthoryear{Shannon, Boyce, Gadwal, and Barnes}{Shannon
  et~al\mbox{.}}{2013}]%
        {shannon2013:tutorial-practices}
\bibfield{author}{\bibinfo{person}{Amy Shannon}, \bibinfo{person}{Acey Boyce},
  \bibinfo{person}{Chitra Gadwal}, {and} \bibinfo{person}{Tiffany Barnes}.}
  \bibinfo{year}{2013}\natexlab{}.
\newblock \showarticletitle{Effective Practices in Game Tutorial Systems}. In
  \bibinfo{booktitle}{{\em Proceedings of the 8th International Conference on
  the Foundations of Digital Games}}.
\newblock


\bibitem[\protect\citeauthoryear{Siorpaes and Hepp}{Siorpaes and Hepp}{2008}]%
        {siorpaes2008:ontogame}
\bibfield{author}{\bibinfo{person}{Katharina Siorpaes} {and}
  \bibinfo{person}{Martin Hepp}.} \bibinfo{year}{2008}\natexlab{}.
\newblock \showarticletitle{Games with a Purpose for the Semantic Web}.
\newblock \bibinfo{journal}{{\em IEEE Intelligent Systems\/}}
  \bibinfo{volume}{23}, \bibinfo{number}{3} (\bibinfo{year}{2008}),
  \bibinfo{pages}{50--60}.
\newblock


\bibitem[\protect\citeauthoryear{Siu, Guzdial, and Riedl}{Siu
  et~al\mbox{.}}{2017}]%
        {siu2017:gwario}
\bibfield{author}{\bibinfo{person}{Kristin Siu}, \bibinfo{person}{Matthew
  Guzdial}, {and} \bibinfo{person}{Mark~O. Riedl}.}
  \bibinfo{year}{2017}\natexlab{}.
\newblock \showarticletitle{Evaluating Singleplayer and Multiplayer in Human
  Computation Games}. In \bibinfo{booktitle}{{\em Proceedings of the 12th
  International Conference on the Foundations of Digital Games}}.
  \bibinfo{pages}{to appear}.
\newblock


\bibitem[\protect\citeauthoryear{Siu and Riedl}{Siu and Riedl}{2016}]%
        {siu2016:hcg-rewards}
\bibfield{author}{\bibinfo{person}{Kristin Siu} {and} \bibinfo{person}{Mark~O.
  Riedl}.} \bibinfo{year}{2016}\natexlab{}.
\newblock \showarticletitle{Reward Systems in Human Computation Games}. In
  \bibinfo{booktitle}{{\em Proceedings of the 2016 Annual Symposium on
  Computer-Human Interaction in Play}} {\em (\bibinfo{series}{CHI PLAY '16})}.
  \bibinfo{publisher}{ACM}, \bibinfo{address}{New York, NY, USA},
  \bibinfo{pages}{266--275}.
\newblock
\showISBNx{978-1-4503-4456-2}
\showDOI{%
\url{http://dx.doi.org/10.1145/2967934.2968083}}


\bibitem[\protect\citeauthoryear{Siu, Zook, and Riedl}{Siu
  et~al\mbox{.}}{2014}]%
        {siu2014:gwap}
\bibfield{author}{\bibinfo{person}{Kristin Siu}, \bibinfo{person}{Alexander
  Zook}, {and} \bibinfo{person}{Mark~O. Riedl}.}
  \bibinfo{year}{2014}\natexlab{}.
\newblock \showarticletitle{Collaboration versus Competition: Design and
  Evaluation of Mechanics for Games with a Purpose}. In
  \bibinfo{booktitle}{{\em Proceedings of the 9th International Conference on
  the Foundations of Digital Games}}.
\newblock


\bibitem[\protect\citeauthoryear{Tuite}{Tuite}{2014}]%
        {tuite2014:gwap-problem}
\bibfield{author}{\bibinfo{person}{Kathleen Tuite}.}
  \bibinfo{year}{2014}\natexlab{}.
\newblock \showarticletitle{{GWAP}s: Games with a Problem}. In
  \bibinfo{booktitle}{{\em Proceedings of the 9th International Conference on
  the Foundations of Digital Games}}.
\newblock


\bibitem[\protect\citeauthoryear{Tuite, Banerjee, Snavely, Popovi\'c, and
  Popovi\'c}{Tuite et~al\mbox{.}}{2015}]%
        {tuite2015:pointcraft}
\bibfield{author}{\bibinfo{person}{Kathleen Tuite}, \bibinfo{person}{Rahul
  Banerjee}, \bibinfo{person}{Noah Snavely}, \bibinfo{person}{Jovan Popovi\'c},
  {and} \bibinfo{person}{Zoran Popovi\'c}.} \bibinfo{year}{2015}\natexlab{}.
\newblock \showarticletitle{PointCraft: Harnessing Players' FPS Skills to
  Interactively Trace Point Clouds in 3D}. In \bibinfo{booktitle}{{\em
  Proceedings of the 10th International Conference on the Foundations of
  Digital Games}}.
\newblock


\bibitem[\protect\citeauthoryear{Tuite, Snavely, Hsiao, Tabing, and
  Popovi\'{c}}{Tuite et~al\mbox{.}}{2011}]%
        {tuite2011:photocity}
\bibfield{author}{\bibinfo{person}{Kathleen Tuite}, \bibinfo{person}{Noah
  Snavely}, \bibinfo{person}{Dun-yu Hsiao}, \bibinfo{person}{Nadine Tabing},
  {and} \bibinfo{person}{Zoran Popovi\'{c}}.} \bibinfo{year}{2011}\natexlab{}.
\newblock \showarticletitle{PhotoCity: Training Experts at Large-scale Image
  Acquisition Through a Competitive Game}. In \bibinfo{booktitle}{{\em
  Proceedings of the SIGCHI Conference on Human Factors in Computing Systems}}
  {\em (\bibinfo{series}{CHI '11})}. \bibinfo{publisher}{ACM},
  \bibinfo{address}{New York, NY, USA}, \bibinfo{pages}{1383--1392}.
\newblock
\showISBNx{978-1-4503-0228-9}
\showDOI{%
\url{http://dx.doi.org/10.1145/1978942.1979146}}


\bibitem[\protect\citeauthoryear{von Ahn and Dabbish}{von Ahn and
  Dabbish}{2004}]%
        {vonahn2004:esp}
\bibfield{author}{\bibinfo{person}{Luis von Ahn} {and} \bibinfo{person}{Laura
  Dabbish}.} \bibinfo{year}{2004}\natexlab{}.
\newblock \showarticletitle{Labeling Images with a Computer Game}. In
  \bibinfo{booktitle}{{\em Proceedings of the SIGCHI Conference on Human
  Factors in Computing Systems}} {\em (\bibinfo{series}{CHI '04})}.
  \bibinfo{publisher}{ACM}, \bibinfo{address}{New York, NY, USA},
  \bibinfo{pages}{319--326}.
\newblock
\showISBNx{1-58113-702-8}
\showDOI{%
\url{http://dx.doi.org/10.1145/985692.985733}}


\bibitem[\protect\citeauthoryear{von Ahn and Dabbish}{von Ahn and
  Dabbish}{2008}]%
        {vonahn2008:gwap-design}
\bibfield{author}{\bibinfo{person}{Luis von Ahn} {and} \bibinfo{person}{Laura
  Dabbish}.} \bibinfo{year}{2008}\natexlab{}.
\newblock \showarticletitle{Designing Games with a Purpose}.
\newblock \bibinfo{journal}{{\it Commun. ACM}} \bibinfo{volume}{51},
  \bibinfo{number}{8} (\bibinfo{year}{2008}), \bibinfo{pages}{58--67}.
\newblock


\bibitem[\protect\citeauthoryear{Wallner and Kriglstein}{Wallner and
  Kriglstein}{2013}]%
        {wallner2013:gameplay-viz-analysis}
\bibfield{author}{\bibinfo{person}{G. Wallner} {and} \bibinfo{person}{S.
  Kriglstein}.} \bibinfo{year}{2013}\natexlab{}.
\newblock \showarticletitle{Visualization-based analysis of gameplay data --- A
  review of literature}.
\newblock \bibinfo{journal}{{\em Entertainment Computing\/}}
  \bibinfo{volume}{4}, \bibinfo{number}{3} (\bibinfo{year}{2013}),
  \bibinfo{pages}{143 -- 155}.
\newblock


\bibitem[\protect\citeauthoryear{Whitlock, McLaughlin, Leidheiser, Gandy, and
  Allaire}{Whitlock et~al\mbox{.}}{2014}]%
        {whitlock2014:flow}
\bibfield{author}{\bibinfo{person}{Laura~A. Whitlock},
  \bibinfo{person}{Anne~Collins McLaughlin}, \bibinfo{person}{William
  Leidheiser}, \bibinfo{person}{Maribeth Gandy}, {and}
  \bibinfo{person}{Jason~C. Allaire}.} \bibinfo{year}{2014}\natexlab{}.
\newblock \showarticletitle{Know Before you Go: Feelings of Flow for Older
  Players Depends on Game and Player Characteristics}. In
  \bibinfo{booktitle}{{\em ACM SIGCHI Annual Symposium on Computer-Human
  Interaction in Play}}.
\newblock


\bibitem[\protect\citeauthoryear{Yee}{Yee}{2006}]%
        {yee2006:motivations}
\bibfield{author}{\bibinfo{person}{Nick Yee}.} \bibinfo{year}{2006}\natexlab{}.
\newblock \showarticletitle{Motivations for play in online games}.
\newblock \bibinfo{journal}{{\em CyberPsychology \& behavior\/}}
  \bibinfo{volume}{9}, \bibinfo{number}{6} (\bibinfo{year}{2006}),
  \bibinfo{pages}{772--775}.
\newblock


\end{thebibliography}
